\documentclass[aps,prx,superscriptaddress,showpacs,twocolumn]{revtex4-2}
\usepackage[left=1.45cm,right=1.45cm,bottom=1.45cm,top=1.45cm,ignoreall]{geometry}
\usepackage[utf8]{inputenc}
\usepackage{graphicx}
\usepackage{amsmath}
\usepackage{amssymb}
\usepackage{xspace}
\usepackage{bm}
\usepackage{color}
\usepackage[squaren,Gray]{SIunits}
\usepackage[hyperindex=true]{hyperref}
\hypersetup{linktocpage,colorlinks=true,citecolor=blue,linkcolor=blue}
\usepackage{empheq}
\usepackage{braket}
\usepackage{physics}
\usepackage{multirow}
\usepackage{array}
\usepackage{makecell}

\begin{document}

\title{Spectrally Robust Photon-Pair Generation in Topological Waveguide Arrays}

\author{A.\ Zecchetto}
\affiliation{Université Paris Cité, CNRS, Laboratoire Matériaux et Phénomènes Quantiques, 75013 Paris, France}

\author{J.-R.~Coudevylle}
\affiliation{Université Paris-Saclay, CNRS, Centre de Nanosciences et de Nanotechnologies, 91120 Palaiseau, France}

\author{M.~Morassi}
\affiliation{Université Paris-Saclay, CNRS, Centre de Nanosciences et de Nanotechnologies, 91120 Palaiseau, France}
	
\author{A.~Lemaître}
\affiliation{Université Paris-Saclay, CNRS, Centre de Nanosciences et de Nanotechnologies, 91120 Palaiseau, France}

\author{M.I.~Amanti}
\affiliation{Université Paris Cité, CNRS, Laboratoire Matériaux et Phénomènes Quantiques, 75013 Paris, France}

\author{S.~Ducci}
\affiliation{Université Paris Cité, CNRS, Laboratoire Matériaux et Phénomènes Quantiques, 75013 Paris, France}

\author{F.~Baboux}
\thanks{\textcolor{blue}{florent.baboux@u-paris.fr}}
\affiliation{Université Paris Cité, CNRS, Laboratoire Matériaux et Phénomènes Quantiques, 75013 Paris, France}
\affiliation{Institut Universitaire de France (IUF)}
	
\makeatletter
\def\Dated@name{} 
\makeatother


\begin{abstract}

Harnessing topological effects offers a route to protect quantum states of light from imperfections, potentially enabling more robust platforms for quantum information processing. 
This capability is particularly relevant for active photonic circuits that generate quantum light directly on-chip. 
However, the impact of topology on the spectral properties of parametric processes---crucial for scalable quantum photonic circuits---remains largely unexplored. 
Here, we address this question by investigating photon-pair generation via spontaneous parametric down-conversion (SPDC) in nonlinear waveguide arrays, combining experiment and theory.
A systematic comparison of uniform, trivial, and topological Su–Schrieffer–Heeger (SSH) arrays reveals that only the topological configuration preserves a stable SPDC resonance spectrum under coupling disorder, with fluctuations in the resonance wavelength reduced by more than an order of magnitude. 
An analytical model captures the origin of this robustness in the band-structure properties of the interacting modes.
Numerical analysis further shows that each SSH array can operate as a robust photon-pair source, enabling stable interference between multiple such sources and the generation of high-dimensional entangled states with high fidelity.
These results establish nonlinear waveguide arrays as a promising platform for harnessing the interplay of topology and nonlinearity in quantum photonic circuits.

\end{abstract}

\maketitle

\section*{Introduction}

Integrated photonic circuits have emerged as a powerful platform for generating and manipulating complex quantum states of light in compact and scalable architectures~\cite{Wang20}. 
Among the various implementations, continuously coupled systems such as waveguide arrays are attracting growing interest for quantum information applications \cite{Bromberg09,Peruzzo10,Solntsev14,Chapman16,Rechtsman16,Grafe16,Tambasco18,BlancoRedondo18,Lahini18,Barral20,Barral21,Hoch22,Raymond24,Raymond25,Chapman24,Yang25}. 
In these structures, photons are continuously transferred between neighboring waveguides through evanescent coupling as they propagate. This enables distributed interference along the full propagation length and gives rise to rich band-structure physics that can be described by lattice Hamiltonians \cite{Perets08}.
This intrinsic connection to condensed-matter models has enabled optical simulations of Anderson localization~\cite{DiGiuseppe13}, Bloch oscillations \cite{Lebugle15}, quantum transport \cite{Caruso16,Biggerstaff16}, and topological phenomena \cite{Ozawa19,Gao24_review,Hashemi25}.

In particular, topological photonic lattices have recently enabled controlled manipulation of quantum states of light in integrated platforms. 
Experiments in passive glass circuits, where quantum states are injected externally, have demonstrated quantum interference of topological single-photon states \cite{Tambasco18}, the robustness of second-order cross-correlation functions in quasi-periodic lattices \cite{Wang19}, or the protection of polarization-entangled states against disorder in Su–Schrieffer–Heeger (SSH) arrays \cite{Wang22}.

In active platforms, where quantum light is generated directly on chip via optical nonlinearities \cite{Leykam15,Yang21,Wang26}, topological effects have also provided tangible benefits.
In silicon-based waveguide arrays exploiting $\chi^{(3)}$ nonlinearity, squeezed states supported by SSH modes have been realized \cite{Ren22}, albeit without probing the effect of disorder.
The topological protection of the spatial profile of biphoton states against disorder, on the other hand, has been convincingly demonstrated in \cite{BlancoRedondo18,Wang19Blanco,Bergamasco19}, further confirming the practical relevance of this approach.

However, these demonstrations primarily concern spatial-mode properties of quantum states.
The protection of \textit{spectral} features of quantum light generation \cite{Mittal18}, such as the resonance spectrum of the underlying nonlinear interaction or the emission spectrum of photon pairs, has not yet been established in waveguide-array platforms.
These features are nevertheless essential for complex photonic circuits, where multiple parametric sources generate quantum states that must subsequently interfere within the circuit \cite{Silverstone14,Wang18,Chapman25}. 
Since these sources are typically pumped by a single laser beam that is split among them, it is important that despite inevitable fabrication imperfections, they all exhibit the same resonance wavelength (with minimal uncertainty), and that they produce photon pairs or heralded single photons with identical emission spectra. 
Achieving such spectral uniformity---a prerequisite for high-visibility quantum interference in complex circuits---raises the question of whether topological protection can be extended to the spectral features of quantum light generation in waveguide arrays.

Here, we address this question theoretically and experimentally by focusing on the resonance spectrum of spontaneous parametric down-conversion (SPDC) in $\chi^{(2)}$ waveguide arrays. We systematically compare different lattice geometries: uniform arrays, arrays featuring a trivial localized mode, and topological SSH arrays in the presence of coupling disorder. While disorder strongly affects the SPDC resonance spectrum in the first two cases, inducing significant fluctuations in both its shape and peak position, the resonance spectrum of topological arrays remains robust up to substantial disorder levels.
Our experiments carried out in AlGaAs waveguide arrays, chosen for their strong $\chi^{(2)}$ nonlinearity enabling efficient on-chip SPDC, are supported by exact numerical simulations and an analytical model that captures these different behaviors and traces them back to the band-structure properties of the interacting modes. 
To our knowledge, these results provide the first experimental demonstration of topological protection of the resonance spectrum of quantum light generation in a one-dimensional system, paving the way for simple and compact on-chip implementations.
A systematic numerical analysis further indicates that this spectral robustness translates into enhanced stability of the generated biphoton states, including greater resilience to fabrication imperfections than conventional single-waveguide sources. 
Each SSH array can therefore act as a robust photon-pair source, allowing stable interference between multiple such sources even in the presence of disorder, opening perspectives for scalable quantum information processing and simulation on-chip.

\begin{figure}[!t]
\centering
\includegraphics[width=0.98\columnwidth]{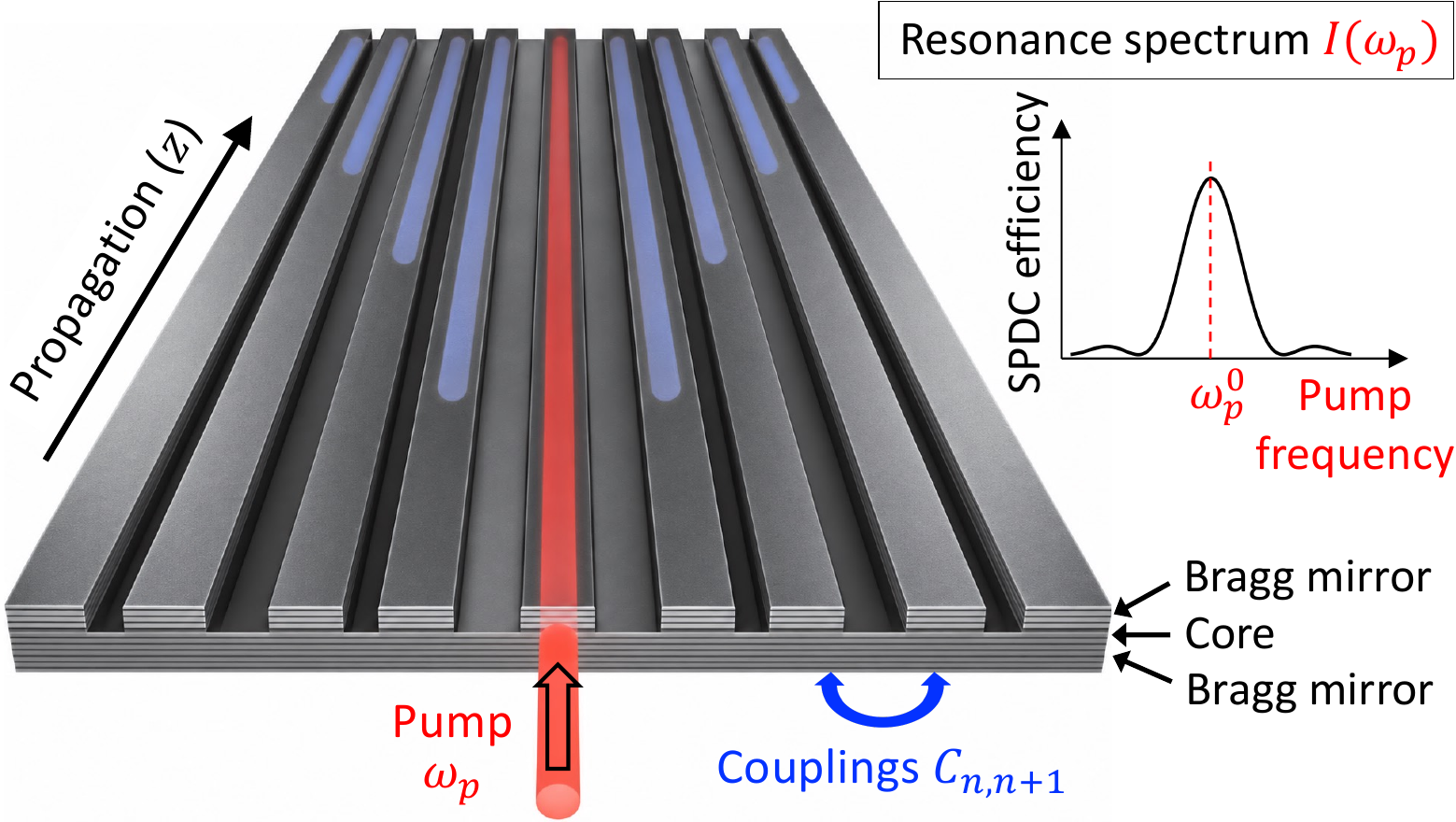}
\caption{Working principle of a $\chi^{(2)}$ nonlinear waveguide array (here implementing the Su–Schrieffer–Heeger model). A monochromatic pump beam at frequency \( \omega_p \) (sketched in red) is injected into the central waveguide, generating pairs of signal and idler photons through spontaneous parametric down-conversion. 
These photons can couple evanescently between adjacent waveguides (as sketched in blue), with coupling constants $C_{n,n+1}$.
The resonance spectrum (inset) quantifies the efficiency of photon-pair generation as a function of the pump frequency and serves here as a figure of merit to probe the stability of the SPDC process against disorder.
}
\label{Fig_Principle}
\end{figure}

\section*{Theoretical framework}

We consider an array of $\chi^{(2)}$ nonlinear waveguides with coupling constants $C_{n,n+1}$ between neighboring waveguides.
A monochromatic pump beam at frequency $\omega_p$ (shown in red in Fig.~\ref{Fig_Principle}) propagates along the array and continuously generates signal and idler photons (frequencies $\omega_s$ and $\omega_i$) by SPDC, which subsequently undergo quantum walks (as sketched in blue) through evanescent coupling between adjacent waveguides.
We assume co-polarized, spectrally degenerate photons, $\omega_s=\omega_i=\omega_p/2$, such that the signal and idler share identical propagation and coupling constants.
The biphoton state at the array output can be written as $\ket{\Psi}=\sum_{n_s,n_i}\Psi_{n_s,n_i}\ket{n_s,n_i}$,
where $\Psi_{n_s,n_i}$ governs the probability amplitude to detect one photon in waveguide $n_s$ and the other in $n_i$. 
It obeys coupled-mode equations along the propagation direction~$z$~\cite{Solntsev12,Grafe12}:
\begin{equation}
	\frac{d\Psi_{n_s,n_i}}{dz}= i\,\mathcal{C}[\Psi] + \gamma \, \delta_{n_s,n_i} \,A_{n_s}(z) e^{i\Delta\beta^{(0)}_{n_s}z},
	\label{EqDiff_short}
\end{equation}
where $\mathcal{C}[\Psi]$ denotes, for compactness, the nearest-neighbor coupling terms \cite{SM}, $\gamma$ quantifies the SPDC efficiency, and the Kronecker delta $\delta_{n_s,n_i}$ enforces that the signal and idler photons are generated in the same waveguide $n_s = n_i \equiv n$.
Here, $A_{n}(z)$ is the pump amplitude in waveguide $n$, and $\Delta\beta^{(0)}_n=\beta^p_n-\beta^s_n-\beta^i_n$ is the single-waveguide phase mismatch, where $\beta^p_n$, $\beta^s_n$, and $\beta^i_n$ are the propagation constants of the pump, signal, and idler fields in waveguide $n$.
The pump field follows analogous coupled-mode equations with reduced coupling constants $C^p_{n,n+1}=\alpha \, C_{n,n+1}$, with $\alpha \ll 1$ since the coupling of the pump is much less efficient due to its halved wavelength~\cite{Solntsev14}. 
Full equations are provided in the Supplemental Material \cite{SM}.

We inject the pump into the central waveguide (labeled $n=0$) and investigate the SPDC resonance spectrum, which quantifies the photon-pair generation efficiency as a function of the pump frequency. We evaluate this spectrum at the output of the central waveguide as $I(\omega_p)=|\Psi_{0,0}(z \!=\!L)|^2$, where $L$ is the array length. Experimentally, this corresponds to coincidence counts recorded while scanning the pump frequency (see inset of Fig.~\ref{Fig_Principle}).

\begin{figure*}[!t]
\centering
\includegraphics[width=0.87\textwidth]{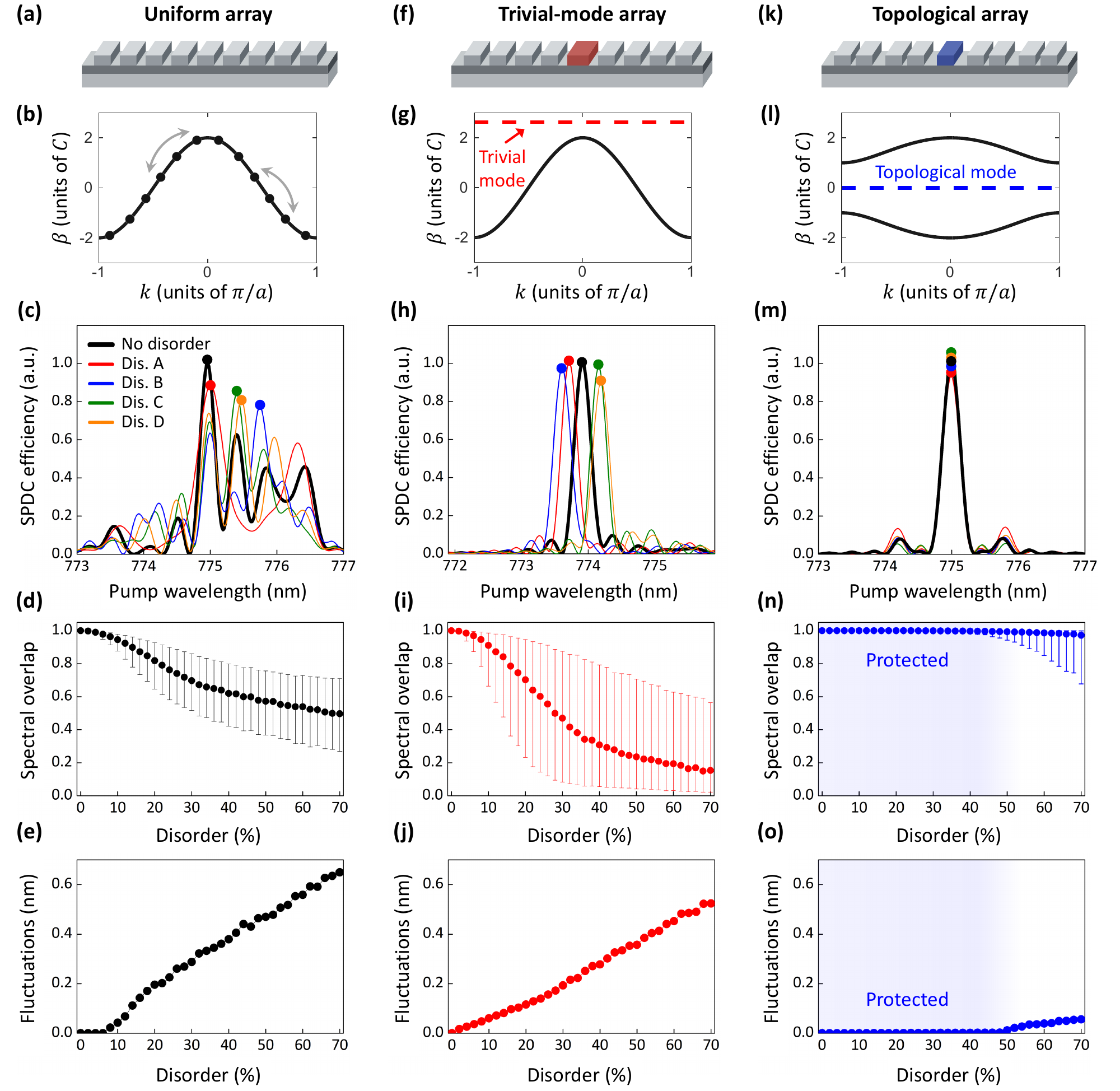}
\caption{
\textbf{First column}: (a) Schematic transverse view and (b) band structure of a uniform waveguide array. 
(c) Calculated SPDC resonance spectra in the absence of disorder (bold black line) and for 4 different disorder realizations (thin colored lines) with disorder strength $\Delta = 40\%$ in the coupling constants, with pump injection and photon-pair collection in the central waveguide.
(d) Median and spread (16th–84th percentiles) of the overlap between each disordered resonance spectrum and the disorder-free spectrum, calculated over 1000 realizations as a function of disorder strength.
(e) Standard deviation of the resonance-peak wavelength as a function of disorder strength. 
\textbf{Second column}, panels (f)-(j): Same quantities for an array featuring a trivial localized mode at its center, resulting from a larger central waveguide (red in panel f).
\textbf{Third column}, panels (k)-(o): Same quantities for a topological array implementing the Su--Schrieffer--Heeger model, with alternating coupling constants $C(1+K)$ and $C(1-K)$, dimerization parameter $K = 0.5$, and a topological interface at the center (blue).
Simulation parameters: all arrays contain 13 waveguides of length $L = 2$~mm, with coupling constant $C = 4$~mm$^{-1}$ and pump coupling parameter $\alpha = 0.1$; the modal dispersions correspond to the AlGaAs waveguides considered in the experiments (see text). In panels (b,g,l), the schematic band structures show the propagation constant $\beta$ relative to that of an isolated waveguide versus the Bloch wave vector $k$ (in units of $\pi/a$, with $a$ the lattice period). The bulk bands (bold lines) correspond to infinite periodic lattices. Black dots in (b) schematically illustrate the bulk eigenmodes of a finite uniform array, while dashed lines in (g) and (l) mark the propagation constants of localized modes.
}
\label{Fig_ResonanceSpectra}
\end{figure*}

We first consider a uniform periodic array, with identical coupling constants $C_{n,n+1} = C$ for all $n$, as sketched in Fig.~\ref{Fig_ResonanceSpectra}a (cross-section view).
Figure~\ref{Fig_ResonanceSpectra}c shows (black bold line) the calculated resonance spectrum of an array of 13 waveguides, using realistic parameters that will be explored experimentally below (see caption for details), with a pump wavelength around 775~nm generating photon pairs in the telecom range.
Since the simplified model of Eq.~\eqref{EqDiff_short} assumes spectrally degenerate signal and idler photons, no spectral integration is involved in the simulations of Fig.~\ref{Fig_ResonanceSpectra}.
In contrast to a single waveguide, which would exhibit a $\mathrm{sinc}^2$ resonance profile, the spectrum displays an enlarged and richer structure. Indeed, inter-waveguide coupling gives rise to a band structure for the propagation constants of both pump and SPDC fields, as illustrated in  Fig.~\ref{Fig_ResonanceSpectra}b in the infinite-lattice limit. The corresponding discrete supermodes of the finite array enable multiple phase-matching channels, producing the observed broad and structured resonance spectrum \cite{SM,Solntsev12,Grafe12}.

We next introduce disorder in the coupling constants to model fabrication imperfections. We assume a uniform distribution of relative amplitude $\Delta$, such that $C_{n,n+1} \in [C(1-\Delta),C(1+\Delta)]$ (which translates to the pump according to $C^p_{n,n+1}=\alpha \, C_{n,n+1}$), and we compute the resonance spectrum for different realizations of disorder $\Delta=40\%$, as shown with thin colored lines in Fig.~\ref{Fig_ResonanceSpectra}c. The spectra are normalized by their area for clarity.
They exhibit strong fluctuations, both in their overall shape and in the position of the resonance peak, marked by dots.
To quantify these variations, we analyze 1000 disorder realizations at each disorder strength and compute the spectral overlap of each resonance spectrum with the disorder-free spectrum, as a measure of spectral stability. Figure~\ref{Fig_ResonanceSpectra}d reports the median spectral overlap and its spread (error bars indicate the variability across realizations). As disorder increases, the median overlap decreases rapidly, while its spread grows. Figure~\ref{Fig_ResonanceSpectra}e further shows that the standard deviation of the resonance-peak position increases monotonically with disorder. Together, these results demonstrate the strong sensitivity of both the resonance shape and the peak position to disorder in uniform waveguide arrays.

We next consider a Su–Schrieffer–Heeger (SSH) array, obtained by alternating short and long spacings between adjacent waveguides \cite{Su79}. A topological interface is introduced at the center by inserting an additional long spacing, as sketched in Fig.~\ref{Fig_ResonanceSpectra}k. This gives rise to a topological mode exponentially localized on the central waveguide (blue), which resides at the center of the band structure (horizontal dotted line in Fig.~\ref{Fig_ResonanceSpectra}l) \cite{BlancoRedondo16}.
Its propagation constant is protected against coupling disorder by chiral symmetry, and isolated from the bulk spectrum by a band gap of width $4KC$ (in the infinite-lattice limit), where $K$ is the dimerization parameter controlling the modulation of the couplings \cite{SM,Asboth16,StJean17,Kailo15}.
Using the same parameters as in Fig.~\ref{Fig_ResonanceSpectra}c, Fig.~\ref{Fig_ResonanceSpectra}m displays the calculated resonance spectra for $K=0.5$, without disorder (bold black line) and for various disorder realizations (thin colored lines) with $\Delta=40\%$. 
In strong contrast with the uniform-array case, the spectra are nearly identical and exhibit a symmetric, single-peak profile. 
Figures~\ref{Fig_ResonanceSpectra}n and \ref{Fig_ResonanceSpectra}o show the statistics of the spectral overlap and of the resonance-peak position as a  function of disorder strength. The spectral overlap remains close to unity, while fluctuations in the resonance-peak position are strongly suppressed compared to the uniform-array case. This spectral stability persists up to disorder strengths of $\Delta \approx 50\%$ (corresponding to the dimerization parameter $K$), around which a gradual crossover toward a more disorder-sensitive regime occurs.

One may question whether this robustness stems from lattice topology itself or simply from spatial localization of the interacting modes, which suppresses transverse propagation in the array and could thereby mitigate the impact of disorder on SPDC generation. 
To address this question, we consider a periodic array, albeit with a slightly wider central waveguide, as sketched in Fig.~\ref{Fig_ResonanceSpectra}f  (hereafter referred to as the ``trivial-mode array''). 
The increased modal index at the center induces a spatially localized mode with an eigenvalue lying above the bulk band (Fig.~\ref{Fig_ResonanceSpectra}g).
This mode is topologically trivial, and its propagation constant is not protected against disorder \cite{StJean17,BlancoRedondo16}.
Figure~\ref{Fig_ResonanceSpectra}h shows the corresponding resonance spectra without disorder (bold black line) and for various realizations of coupling disorder (thin colored lines) with $\Delta=40\%$.
While the spectrum remains single-peaked, with a shape comparable to that of the topological array, the peak position fluctuates strongly with disorder (Fig.~\ref{Fig_ResonanceSpectra}j), leading to a rapid degradation of the spectral overlap as disorder increases (Fig.~\ref{Fig_ResonanceSpectra}i).
This comparison demonstrates that spatial localization alone is insufficient to ensure robustness of the parametric resonance against disorder, highlighting the protective role of the SSH topology.

To complement the numerical simulations and elucidate the origin of these contrasting behaviors, we now introduce a simple analytical model capturing the essential physics. 
We express the biphoton state in the basis of spatial supermodes (lattice eigenmodes) as
\begin{equation}
	\ket{\Psi}
	=
	 \sum_{j} 
	A_{m_p}\gamma_j  \, L  \, 
	e^{i\Delta\beta_j L/2}
	\,\mathrm{sinc}\!\left(\Delta\beta_j L/2\right)
	\hat a_{m_s}^\dagger
	\hat a_{m_i}^\dagger
	\ket{0},
	\label{eq:multimode}
\end{equation}
where the sum runs over all combinations of pump, signal, and idler supermodes, $j=\{m_p,m_s,m_i\}$. 
Here, $\hat a_{m_s}^\dagger$ ($\hat a_{m_i}^\dagger$) creates a signal (idler) photon in supermode $m_s$ ($m_i$), 
$A_{m_p}$ is the pump amplitude in supermode $m_p$, 
and $\gamma_j$ denotes the nonlinear overlap integral of the three interacting modes for the combination $j$. 
The corresponding phase mismatch reads
\begin{equation}
	\Delta \beta_j(\omega_p) = \beta^p_{m_p} - \beta^s_{m_s} - \beta^i_{m_i}.
\end{equation}

In a uniform array, all supermodes are \textit{bulk modes} (modes delocalized across the array and belonging to the propagation band) in the absence of disorder. 
Pumping the central waveguide excites multiple pump supermodes, which can then down-convert into various combinations of signal and idler bulk modes, as schematically illustrated by gray arrows in Fig.~\ref{Fig_ResonanceSpectra}b. 
Disorder alters both the propagation constants and spatial profiles of these modes, thereby modifying the phase-matching frequencies and overlap integrals, respectively.
This naturally leads to strong fluctuations in both the shape and peak position of the resonance spectrum, consistent with the numerical findings of Fig.~\ref{Fig_ResonanceSpectra} (first column).

In contrast, both SSH and trivial-mode arrays support a localized supermode that is efficiently excited by central pumping. 
The dominant SPDC interaction then involves localized pump, signal, and idler modes, yielding a narrow, single-peak spectrum characteristic of a single-mode process.
The crucial difference lies in the behavior of the propagation constant of this mode: in the SSH array it is protected by the underlying chiral symmetry (both for the pump and the SPDC fields), whereas no such protection exists in the trivial configuration. 
As a result, the phase-matching condition associated with this mode remains stable in the SSH array, while it fluctuates with disorder in the trivial case.
Consequently, only the topological array preserves both the shape and the central wavelength of its resonance spectrum, up to significant disorder levels.
Beyond this protection against coupling disorder, the SSH array also provides enhanced resilience to waveguide-width fluctuations compared with an isolated waveguide, as will be discussed below [Fig.~\ref{Fig_Biphoton}f].

The supermode model of Eq.~\eqref{eq:multimode} also explains the onset of small disorder-induced fluctuations once the disorder strength becomes comparable to the dimerization parameter $K$ (here $50\%$), as seen in Figs.~\ref{Fig_ResonanceSpectra}n,o.
At this point, disorder can locally invert the alternation between short and long couplings and significantly modify the modal structure: the original topological supermode may shift spatially or broaden, while some bulk modes may localize near the central (pumped) waveguide and approach the propagation constant of the topological mode.
As a result, additional channels involving pump and/or down-converted photons in disorder-sensitive bulk modes progressively gain importance~\cite{Bergamasco19}. Their relative weight is governed by the nonlinear overlap integrals  $\gamma_j$ and phase-matching $\mathrm{sinc}$ factors appearing in Eq.~\eqref{eq:multimode}, both of which evolve as disorder modifies the modal structure. This produces a crossover from a protected to a more disorder-sensitive nonlinear response beyond $\Delta \approx K$. The resulting fluctuations nevertheless remain much smaller than in non-topological arrays, because the topological channel still dominates the SPDC interaction and thereby ensures effective stabilization of the resonance against coupling disorder.

\begin{figure}[!t]
\centering
\includegraphics[width=8.2cm]{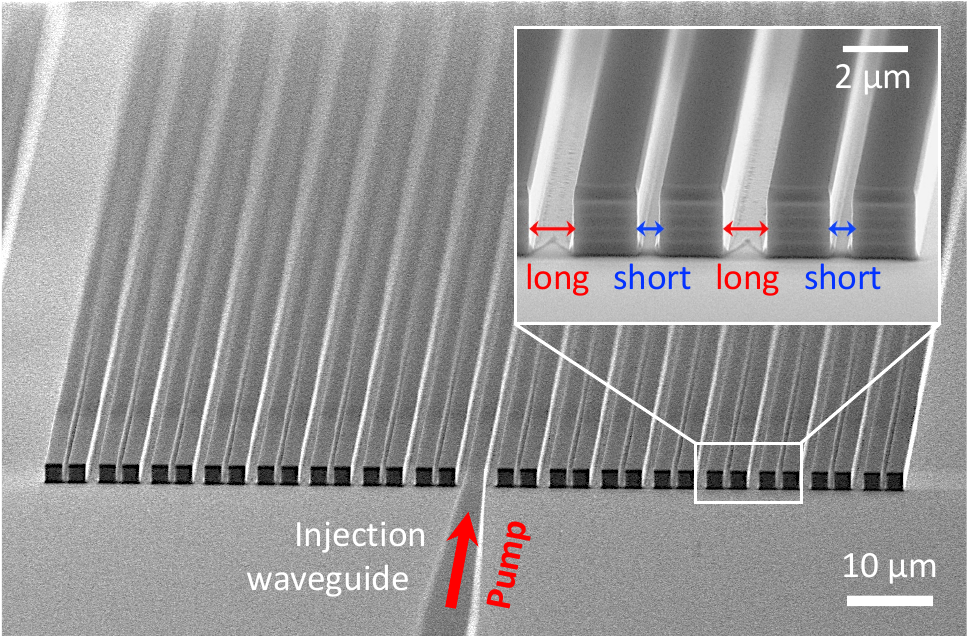}
\caption{SEM image of an AlGaAs topological SSH array (here comprising 31 waveguides) with an injection waveguide for the pump beam.
The inset shows a close-up of the alternating short (870~nm) and long (1700~nm) spacings between adjacent waveguides, resulting in alternating high and low coupling constants. 
}
\label{Fig_Setup}
\end{figure}

\begin{figure*}[!t]
\centering
\includegraphics[width=17.3cm]{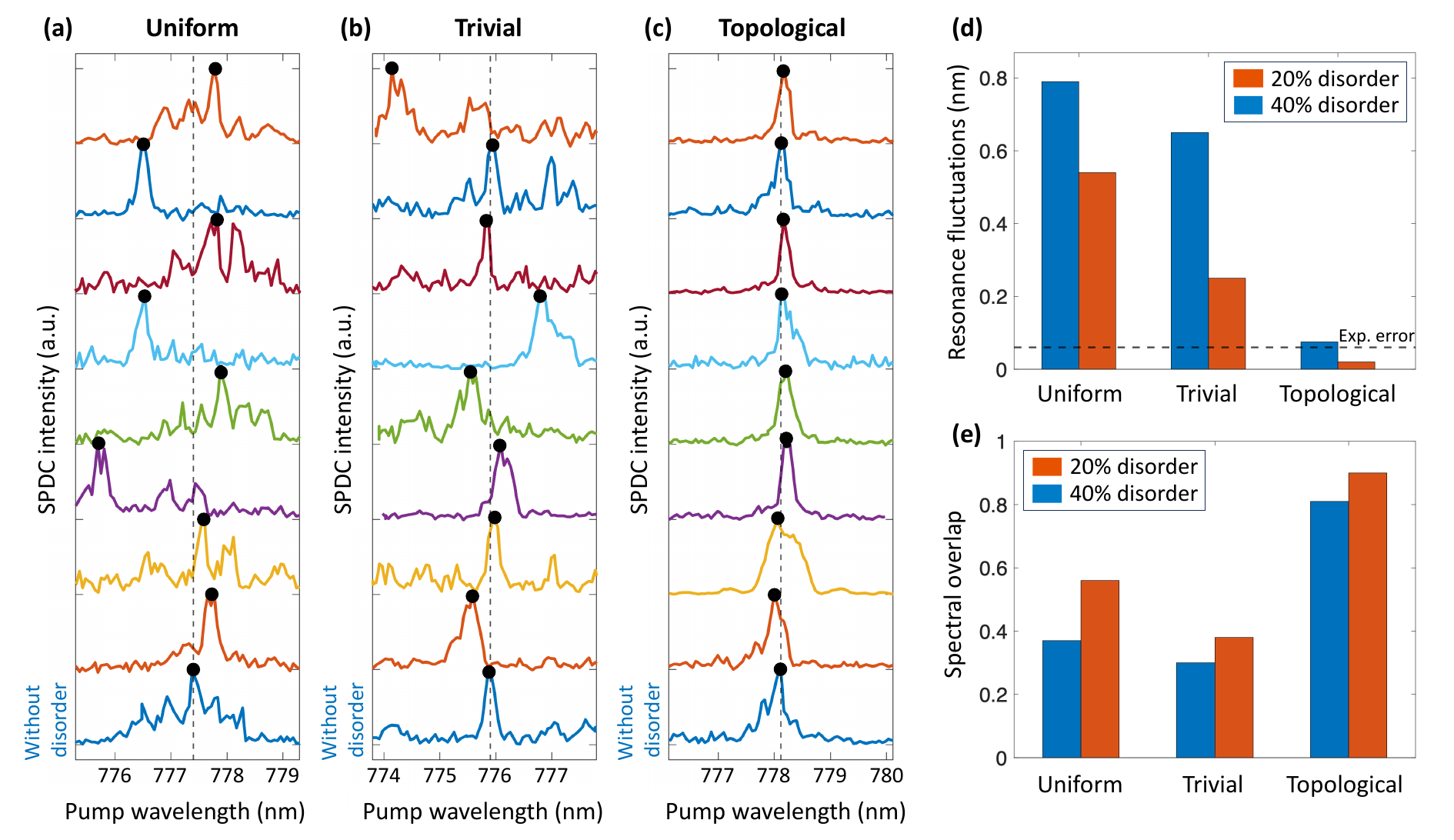}
\caption{
\textbf{(a)} Measured SPDC resonance spectra in uniform waveguide arrays without disorder (bottom trace) and for various random disorder realizations with amplitude \( \Delta = 40\% \). 
\textbf{(b)} Same measurement for arrays featuring a trivial localized mode at the center, and \textbf{(c)} for topological SSH arrays. 
All spectra are normalized to unity and vertically offset for clarity. 
\textbf{(d)}~Experimental fluctuations (standard deviation) of the resonance-peak wavelength for the three types of arrays at coupling disorder strengths \( \Delta = 40\% \) (blue bars) and \( \Delta = 20\% \) (red bars). 
\textbf{(e)} Median pairwise overlap between the SPDC resonance spectra, under the same conditions as in (d).
}
\label{Fig_Experiments}
\end{figure*}

\section*{Experiments}

We now turn to the experimental verification of these predictions.
The three lattice geometries investigated above were fabricated using AlGaAs, which combines a strong $\chi^{(2)}$ nonlinearity and mature nanofabrication technology \cite{Dietrich16}. Each array consists of 13 waveguides with a length of $L = 2$~mm and a width of 2~$\mu$m. 
The arrays were patterned by electron-beam lithography with a high-resolution HSQ resist, followed by inductively coupled plasma (ICP) etching. 
As sketched in Fig.~\ref{Fig_Principle}, the  epitaxial structure comprises a 6-period Al$_{0.8}$Ga$_{0.2}$As/Al$_{0.25}$Ga$_{0.75}$As lower Bragg reflector, a 350~nm Al$_{0.45}$Ga$_{0.55}$As core, and a 4-period Al$_{0.8}$Ga$_{0.2}$As/Al$_{0.25}$Ga$_{0.75}$As upper Bragg reflector.
The two Bragg mirrors provide vertical photonic band-gap confinement for the pump beam in the near-infrared, while ensuring total internal reflection for the down-converted photons in the telecom range. 
As a result, the pump and SPDC modes exhibit distinct dispersion relations, enabling the single-waveguide phase-matching condition to be satisfied within the spectral range of interest \cite{Helmy11,Horn12,Baboux23,Raymond24,Raymond25}.
The central waveguide comprises enlarged input and output sections (6~$\mu$m wide) connected to the array through tapers, as shown in the SEM image in Fig.~\ref{Fig_Setup}. 
These enlarged sections shift the local nonlinear resonance wavelength, ensuring that photon pairs are generated exclusively within the central array region.

For the uniform arrays, the inter-waveguide distance, defined edge-to-edge, is 1100~nm, and the etching depth reaches the first layer above the core, yielding a mean simulated coupling constant $C = 4$~mm$^{-1}$ ($3.5$~mm$^{-1}$ for TE and $4.5$~mm$^{-1}$ for TM polarization) for the SPDC photons and a pump coupling parameter $\alpha = C_p/C \simeq 0.1$, in good agreement with direct measurements based on laser-beam propagation \cite{SM}.
For the topological SSH array, a dimerization $K = 0.5$, centered around the same mean coupling as in the uniform array, is obtained by alternating 870~nm and 1700~nm inter-waveguide distances, as shown in the close-up SEM image of Fig.~\ref{Fig_Setup}.
Finally, for the trivial-mode array, the central waveguide width is \(2.1~\mu\mathrm{m}\) and the inter-waveguide distance is \(1000~\mathrm{nm}\), yielding a defect amplitude identical to that used in the simulations and chosen to reproduce a localization length similar to that of the SSH array. Typical propagation losses of the fabricated structures are about \(1~\mathrm{cm}^{-1}\) at telecom wavelengths.

To investigate the effect of disorder, we intentionally varied the inter-waveguide distances around their nominal values and fabricated a series of samples corresponding to different random disorder realizations  spanning a representative statistical ensemble.
The experiments are then performed by injecting a tunable TE-polarized continuous-wave pump laser into the central waveguide of each array using a microscope objective, with a coupled power of 1~mW.
Orthogonally polarized signal and idler photons are generated via type-II SPDC and collected from the central waveguide with a second microscope objective. 
The SPDC photons are then coupled into an optical fiber, split by a fiber polarizing beamsplitter, and detected by superconducting nanowire single-photon detectors (SNSPDs) after passing through a high-pass filter with a cutoff wavelength of 1500~nm. The filter transmission, together with energy conservation, defines an effective detection bandwidth of $\simeq 100$~nm for the SPDC photons.

Resonance spectra are obtained by tuning the wavelength of the pump laser (Toptica DL pro) in 60~pm steps using a motorized scanning routine, while monitoring the coincidence counts collected from the central waveguide.
Figure~\ref{Fig_Experiments}a shows the SPDC resonance spectra, defined as the signal--idler raw coincidence counts measured as a function of pump wavelength, in a uniform waveguide array without disorder (bottom trace) and for several disorder realizations with amplitude $\Delta = 40\%$. 
All spectra are normalized to unity and vertically offset for clarity. 
The pump-wavelength window shown in Figs.~\ref{Fig_Experiments}a--c was chosen to focus on the main resonance region (it does not represent a spectral boundary of the SPDC emission).
Strong fluctuations are observed in both the spectral shape and the resonance-peak position (marked by black dots).  
The standard deviation of the peak position is 0.79~nm, and the median overlap between the spectra is 0.37.
We checked that the raw spectra exhibit comparable peak coincidence counts ($\sim 200$) across disorder realizations and lattice geometries, indicating sufficient signal levels and comparable Poissonian counting uncertainties. We further verified that either subtracting the measured accidental background or normalizing the coincidence counts by it does not change the extracted resonance-peak positions within the measurement resolution, confirming the consistency of their experimental determination.

We next investigate arrays supporting a trivial localized mode at the center.
The corresponding SPDC resonance spectra, shown in Fig.~\ref{Fig_Experiments}b, exhibit a generally more regular shape, but the resonance-peak positions still display large fluctuations.
Here, the standard deviation of the peak position is 0.65~nm, and the median spectral overlap is 0.30.

Finally, Fig.~\ref{Fig_Experiments}c presents resonance spectra measured in the topological SSH arrays.  
In this case, a much-improved stability is observed, both in the general spectral shape and in the resonance-peak position.  
The standard deviation of the peak position is reduced to 0.075~nm (which approaches the 0.06~nm resolution of the pump laser tuning), while the median overlap increases to 0.81. 
At the resonance peak, the generation rate in the SSH arrays is on the order of $10^5$ pairs.s$^{-1}$.

The experimental results for the three types of arrays are summarized in Fig.~\ref{Fig_Experiments}d (resonance-peak fluctuations) and Fig.~\ref{Fig_Experiments}e (spectral overlap) with blue vertical bars.
To investigate the scaling of spectral fluctuations with the disorder strength, we repeated the same measurements on a second series of samples with a reduced disorder amplitude of 20\%. The corresponding results are shown in Figs.~\ref{Fig_Experiments}d,e with red vertical bars.
In good qualitative agreement with the simulations of Fig.~\ref{Fig_ResonanceSpectra}, reducing the disorder amplitude from 40\% to 20\% leads, in all three cases, to a decrease in resonance-peak fluctuations together with an increase in spectral overlap. 
For the topological arrays, the resonance fluctuations are reduced to about 0.02~nm, while they are respectively 12 and 27 times larger in the trivial-mode and uniform arrays.
In addition, the spectral overlap rises to 0.9 in the SSH arrays, highlighting the strongly reduced influence of disorder on the entire resonance spectrum.
At 40\% disorder, the slight residual resonance-peak fluctuations (0.075~nm) likely arise from the proximity to the point where resonance stabilization begins to weaken ($\Delta \approx K$), as shown in Figs.~\ref{Fig_ResonanceSpectra}n and~o.
A slightly lower effective dimerization parameter $K$ in the fabricated structure may therefore already place the system at the onset of this regime.

We thus find good qualitative agreement with the simulations of Fig.~\ref{Fig_ResonanceSpectra}, based on Eq.~\eqref{EqDiff_short}, which assumes identical polarization and perfect spectral degeneracy of the SPDC photons and captures the essential physics within a minimal framework.
A more quantitative description can be obtained by extending this formalism to account for type-II polarizations, the finite spectral bandwidth of the photons, and the associated polarization and frequency dependence of the coupling constants (see Supp.~Mat.~\cite{SM}).
This extended analysis confirms that the stabilization of the SSH resonance spectrum persists beyond the idealized degenerate model, remaining effective when polarization and spectral non-degeneracies are included and across all detection bandwidths considered. By contrast, the simplified model (Fig.~\ref{Fig_ResonanceSpectra}) shows that non-topological arrays already exhibit significant resonance fluctuations in the \textit{a priori} most favorable limit of co-polarized photons degenerate in frequency.
Overall, these results establish that the topological stabilization of the resonance is robust to the detailed dynamics of the parametric process.
Numerical simulations further indicate that this protection is not specific to the local coincidence observable chosen for our experiment. It extends to first-order intensity observables, including the total single-count signal, showing that the resonance stabilization extends to both local and global observables~\cite{SM}.

\section*{Discussion and perspectives}

\begin{figure*}[!t]
\centering
\includegraphics[width=0.85\textwidth]{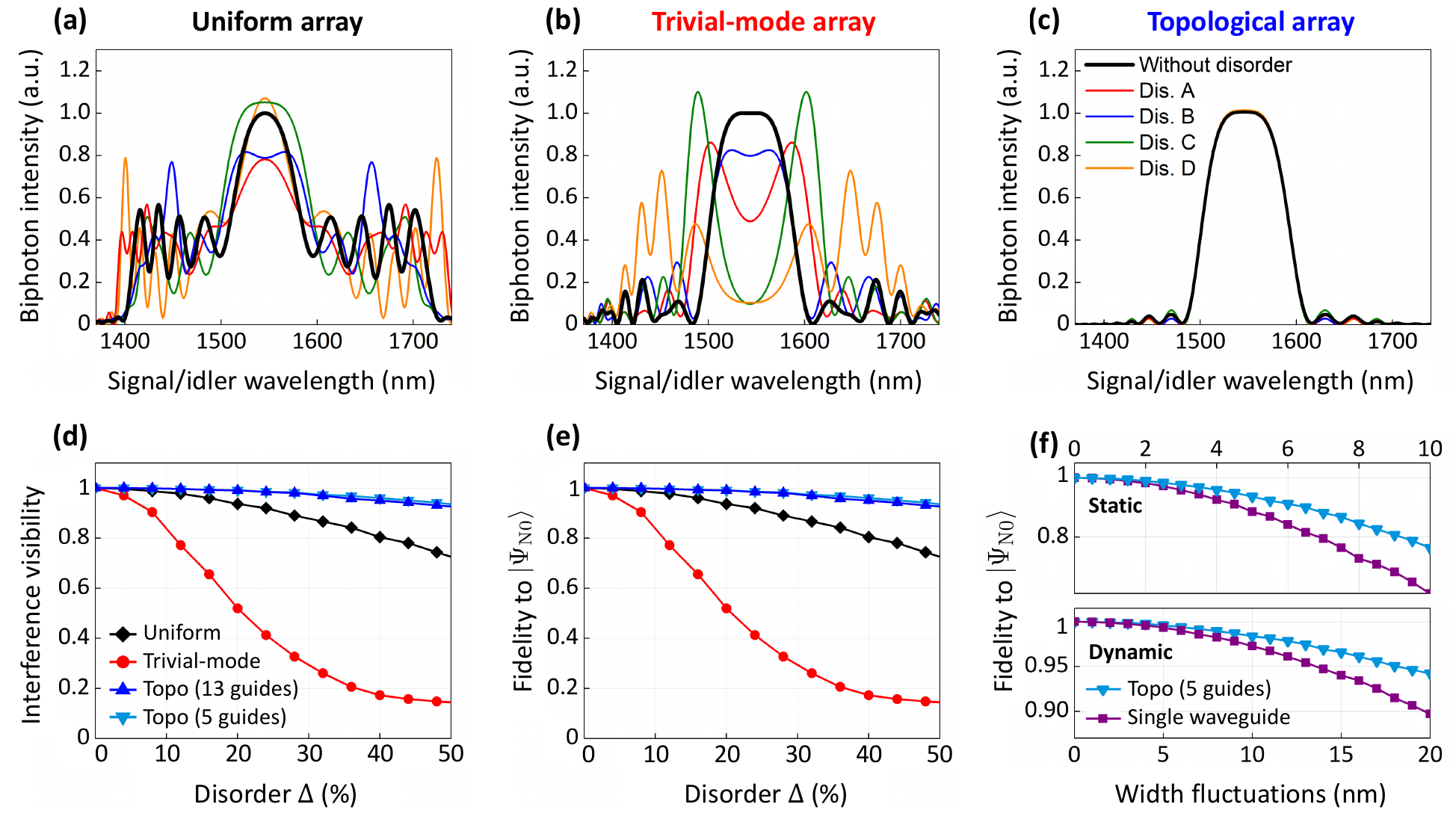}
\caption{
\textbf{(a)} Joint spectral intensity $|\phi(\omega)|^2$ of photon pairs emitted from the central waveguide of a uniform waveguide array, calculated in the absence of disorder (bold black line) and for 4 different disorder realizations (thin colored lines) with disorder strength $\Delta = 40\%$ in the coupling constants. The pump wavelength is tuned to the maximum of the resonance spectrum in the absence of disorder.
\textbf{(b)} Joint spectral intensity calculated for trivial-mode arrays and \textbf{(c)} topological SSH arrays.
\textbf{(d)} Mean interference visibility $\overline{V}=\langle V_{n,n'}\rangle$ between pairs of sources [Eq.~\eqref{eq:visibility}], calculated as a function of coupling disorder, for uniform (black), trivial-mode (red) and topological SSH arrays with 13 waveguides (blue) or 5 waveguides (light blue).
\textbf{(e)} Mean fidelity $\bar{\mathcal{F}}$ to an ideal broadband (100 nm) maximally path-entangled state prepared by coherently pumping 10 biphoton sources, calculated as a function of coupling disorder (same color code).
\textbf{(f)} Mean fidelity (as in (e)) versus waveguide-width fluctuations, for 5-waveguide SSH arrays (light blue) and single-waveguide sources (purple). Top: static disorder. Bottom: longitudinally varying disorder (see text).
}
\label{Fig_Biphoton}
\end{figure*}

Having experimentally demonstrated the robustness of the SPDC resonance spectrum of SSH arrays to coupling disorder, an important question is whether this robustness also extends to the biphoton emission spectrum~\cite{Mittal18}.
Indeed, ensuring that independent parametric sources emit quantum states with identical spectra is essential for achieving high-visibility quantum interference~\cite{Silverstone14,Spring17,Chapman25}. Such interference underpins many quantum information tasks, from quantum computing and simulation to the generation of high-dimensional entangled states~\cite{Brod19,Wang20}.

We investigate how disorder affects the biphoton spectrum in the different arrays considered above through numerical simulations.
Figure~\ref{Fig_Biphoton}a shows the calculated joint spectral intensity ($\vert \phi(\omega)\vert^2$) of photon pairs emitted from the central waveguide of a uniform array (using the same parameters as in Fig.~\ref{Fig_ResonanceSpectra}), in the absence of disorder (bold black line) and for various random disorder realizations (thin colored lines) with $\Delta = 40\%$. Owing to strict energy conservation under CW pumping, the joint spectral amplitude $\phi$ depends on a single frequency variable~$\omega$, such that $\phi(\omega_s,\omega_i) = \phi(\omega_s,\omega_p-\omega_s)\equiv \phi(\omega)$. The pump wavelength is tuned to the maximum of the resonance spectrum of the disorder-free array.
Similarly, Figs.~\ref{Fig_Biphoton}b and c show the joint spectra calculated in trivial-mode arrays and SSH arrays, respectively.
While large spectral fluctuations are observed for uniform and trivial-mode arrays, the biphoton spectra of the topological arrays remain remarkably stable under disorder.

Here, each array is pumped and collected through its central waveguide, and therefore behaves as an effective photon-pair source emitting into a single spatial mode (while its emission spectrum remains shaped by the full lattice modal structure). This raises the question of how stable the interference between such effective sources remains in the presence of disorder.
To address this question, we first consider two independent effective sources pumped by the same continuous-wave laser. Their outputs interfere on a balanced beamsplitter, forming the time-reversed analogue of the Hong--Ou--Mandel experiment~\cite{BlancoRedondo18,Chen07,Silverstone14,Spring17,Chapman25,Fabre20,Fabre22,Meskine24,Descamps26}. Perfect antibunching occurs when the two sources emit biphotons with identical spectra and equal generation efficiencies, while spectral or efficiency mismatches reduce the interference visibility. For two sources $n$ and $n'$, the visibility can be written as~\cite{SM}
\begin{equation}
	V_{n,n'}=
	2\sqrt{P_n P_{n'}}
	\;\Re\!\left[
	\int d\omega\,
	\phi_n^*(\omega)\,\phi_{n'}(\omega)
	\right],
	\label{eq:visibility}
\end{equation}
where $P_n$ and $P_{n'}$ denote the probabilities for generating the photon pair in sources $n$ and $n'$ (under identical pump power), with $P_n+P_{n'}=1$, $\phi_{n}(\omega)$ and $\phi_{n'}(\omega)$ are the normalized joint spectral amplitudes,  and $\Re$ denotes the real part.
The interference visibility therefore provides a comprehensive figure of merit, capturing both resonance stability (via the generation probabilities) and the joint spectral amplitude of the generated photons.

We compute the biphoton states for many realizations of coupling disorder (1000), each corresponding to a distinct source, and evaluate the mean visibility \(\overline{V}=\langle V_{n,n'}\rangle\) over all pairs of sources.
Figure~\ref{Fig_Biphoton}d shows the evolution of $\overline{V}$ as a function of the disorder strength. For the topological SSH array (blue), a high interference visibility is preserved, remaining above 95\% up to 40\% disorder. For comparison, results for uniform and trivial-mode arrays are also shown (black and red), and degrade more rapidly as disorder increases. This behavior, which is particularly pronounced for the trivial-mode array, further demonstrates that mere spatial localization of the SPDC modes is not sufficient to ensure robustness against disorder, highlighting the protective role of the SSH configuration. The same figure also shows results obtained for an SSH array comprising only 5 waveguides (light blue). Interestingly, the performance is very similar to that of the 13-waveguide SSH array, showing that surrounding the central waveguide with just two neighbors on each side, forming a minimal SSH unit, is already sufficient to achieve efficient protection, while minimizing the footprint on the chip.

Beyond pairwise interference, large-scale quantum photonic circuits require many independent sources to emit mutually indistinguishable quantum states. We therefore examine a more demanding scenario in which $N$ parametric sources are pumped simultaneously by the same laser.
In the ideal case, all sources emit identical biphoton spectra $\phi_n(\omega)\equiv\phi_0(\omega)$ and have equal generation efficiencies, such that the resulting state is $ \ket{\Psi_{\mathrm{N0}}} = \frac{1}{\sqrt{N}} \sum_{n=1}^{N} \ket{\phi_0}_n $.
It constitutes a broadband, maximally path-entangled state, namely a coherent superposition of indistinguishable biphotons across $N$ different spatial modes.
To quantify deviations from this ideal situation, we evaluate the fidelity $\mathcal{F}$ between the generated state and the ideal one \cite{SM}.
Figure~\ref{Fig_Biphoton}e shows, for path-entangled states constructed with $N=10$ sources over a 100 nm spectral bandwidth, the mean fidelity $\bar{\mathcal{F}}$ as a function of the disorder strength. A trend similar to that observed for the two-source interference visibility is found. The topological SSH array (with 13 or 5 waveguides) outperforms the non-topological arrays, with a mean fidelity to the ideal state $\ket{\Psi_{\mathrm{N0}}}$ above 96\% up to 40\% coupling disorder.

Finally, we examine the impact of waveguide width fluctuations, which modify the propagation constants $\beta$ of the SPDC and pump modes. In the SSH model, such perturbations break chiral symmetry and therefore lift exact topological protection \cite{Asboth16,Scollon20}.
We here benchmark topological arrays against conventional single-waveguide sources, which provide a natural reference system for such disorder.
For a conservative comparison, we assume that width fluctuations in the topological arrays also affect the corresponding inter-waveguide distances, thereby inducing disorder in the coupling constants as well.
Figure~\ref{Fig_Biphoton}f compares the mean fidelity $\bar{\mathcal{F}}$ to the maximally path-entangled state $\ket{\Psi_{\mathrm{N0}}}$ as a function of width fluctuations, for SSH arrays comprising five waveguides (light blue) and for single waveguides (purple).
The upper panel corresponds to static width inhomogeneities (constant along $z$), while the lower panel considers longitudinally varying fluctuations (with a correlation length of $100 \, \mu$m, representative e.g.~of stitching field errors in the electron-beam lithography process).
Interestingly, SSH arrays consistently outperform single-waveguide sources, for both static and dynamic disorder.
This enhanced robustness arises from the reduced sensitivity of the supermode propagation constant to local width fluctuations. Indeed, in a single waveguide, phase matching directly depends on the local propagation constant and is therefore fully exposed to disorder, whereas the collective nature of the topological SSH supermode averages over individual waveguide perturbations, thereby mitigating their impact.
This mechanism relies on inter-waveguide coupling, which could in principle introduce an additional sensitivity to disorder that is absent in a single waveguide. However, the SSH configuration is precisely resilient to these perturbations: the off-diagonal disorder associated with inter-waveguide distance fluctuations was included in the simulations of Fig.~\ref{Fig_Biphoton}f for completeness, but we verified that it does not affect the fidelity (within numerical precision). The topological system can therefore exploit the self-averaging of diagonal perturbations provided by the collective supermode while remaining protected against coupling disorder, making it a well-suited platform for photon-pair generation with enhanced robustness against various types of fabrication imperfections.

In conclusion, by systematically comparing uniform, trivial-mode, and topological SSH nonlinear waveguide arrays, we have experimentally established the topological robustness of the SPDC resonance spectrum in SSH arrays and identified its origin in the stability of the underlying interacting modes. 
To our knowledge, this provides the first experimental evidence of such protection in a one-dimensional photonic system, opening the way for simple and compact on-chip implementations.
Numerical analysis further predicts that this robustness extends to the spectral properties of the emitted biphoton states, highlighting the potential of topological waveguide arrays to realize robust photon-pair sources enabling high-visibility quantum interference and the generation of broadband high-dimensional entangled states with high fidelity.

The present analysis also clarifies the conditions under which these topological advantages become experimentally observable, and provides a concrete basis for their implementation in suitably designed devices supporting multi-source operation and readout.
As we have shown, efficient protection already arises in small SSH units with as few as five waveguides. Compact architectures could therefore be envisioned, in which several topological interfaces (as sketched in blue in Fig.~\ref{Fig_ResonanceSpectra}k) are periodically embedded along a single chain~\cite{Bergamasco21}. 
For sufficiently large dimerization, the interface states have negligible mutual overlap while each remains embedded in a local SSH environment, allowing the corresponding interfaces to operate as independent and robust photon-pair sources and thereby forming a compact multi-source architecture.
Beyond this specific implementation, our approach could be extended to emerging material platforms that do not yet benefit from mature fabrication \cite{Wang20,Elshaari20}, thereby mitigating imperfections and leveraging topology as a practical route toward resilient quantum photonic circuits.

Looking further ahead, the versatility of the present platform also paves the way for extending topological effects into synthetic dimensions, where internal degrees of freedom, such as frequency modes, emulate additional propagation directions \cite{Yuan18}. Indeed, the strong electro-optic response of AlGaAs \cite{Wang14,Dietrich16} in a traveling-wave geometry could provide controlled motion along a synthetic frequency axis in addition to the spatial dimension~\cite{Piccioli22,Wu22}. This would open the way to protected frequency conversion or hybrid entanglement across time, space, and frequency degrees of freedom~\cite{Piccioli22}, enabling quantum information processing across multiple dimensions.
Nonlinear waveguide arrays thus offer a versatile platform to explore topological effects in the quantum regime and design novel, topology-driven optical functionalities.

\section*{Acknowledgments}

We thank A.~Raymond, L.~Lazzari, F.~Raineri, E.~Descamps and P. Milman for fruitful discussions and P.~Filloux, B.~Janvier and M.~Nicolas for technical support. We acknowledge support from the Ville de Paris Emergence program (\textsc{LATTICE} project), Région Île-de-France in the framework of the DIM QuanTiP (\textsc{Q-LAT} project), IdEx Université Paris Cité (ANR-18-IDEX-0001), the Plan France 2030 through PEPR project NISQ2LSQ (ANR-22-PETQ-0006), and the "Investissement d'Avenir" program of the French Government (ANR‐22‐CMAS-0001, QuanTEdu-France). We acknowledge the use of cleanroom facilities from the “Groupement des Plateformes de Micro et Nanotechnologies de Paris” and support from Université Paris Cité, CNRS, RENATECH+ and RENATECH project (ANR-22-PEEL-0015) from PEPR Electronique for micro- and nanofabrication.


%


\begin{thebibliography}{63}%
	\makeatletter
	\providecommand \@ifxundefined [1]{%
		\@ifx{#1\undefined}
	}%
	\providecommand \@ifnum [1]{%
		\ifnum #1\expandafter \@firstoftwo
		\else \expandafter \@secondoftwo
		\fi
	}%
	\providecommand \@ifx [1]{%
		\ifx #1\expandafter \@firstoftwo
		\else \expandafter \@secondoftwo
		\fi
	}%
	\providecommand \natexlab [1]{#1}%
	\providecommand \enquote  [1]{``#1''}%
	\providecommand \bibnamefont  [1]{#1}%
	\providecommand \bibfnamefont [1]{#1}%
	\providecommand \citenamefont [1]{#1}%
	\providecommand \href@noop [0]{\@secondoftwo}%
	\providecommand \href [0]{\begingroup \@sanitize@url \@href}%
	\providecommand \@href[1]{\@@startlink{#1}\@@href}%
	\providecommand \@@href[1]{\endgroup#1\@@endlink}%
	\providecommand \@sanitize@url [0]{\catcode `\\12\catcode `\$12\catcode
		`\&12\catcode `\#12\catcode `\^12\catcode `\_12\catcode `\%12\relax}%
	\providecommand \@@startlink[1]{}%
	\providecommand \@@endlink[0]{}%
	\providecommand \url  [0]{\begingroup\@sanitize@url \@url }%
	\providecommand \@url [1]{\endgroup\@href {#1}{\urlprefix }}%
	\providecommand \urlprefix  [0]{URL }%
	\providecommand \Eprint [0]{\href }%
	\providecommand \doibase [0]{https://doi.org/}%
	\providecommand \selectlanguage [0]{\@gobble}%
	\providecommand \bibinfo  [0]{\@secondoftwo}%
	\providecommand \bibfield  [0]{\@secondoftwo}%
	\providecommand \translation [1]{[#1]}%
	\providecommand \BibitemOpen [0]{}%
	\providecommand \bibitemStop [0]{}%
	\providecommand \bibitemNoStop [0]{.\EOS\space}%
	\providecommand \EOS [0]{\spacefactor3000\relax}%
	\providecommand \BibitemShut  [1]{\csname bibitem#1\endcsname}%
	\let\auto@bib@innerbib\@empty
	\bibitem [{\citenamefont {Wang}\ \emph {et~al.}(2020)\citenamefont {Wang},
		\citenamefont {Sciarrino}, \citenamefont {Laing},\ and\ \citenamefont
		{Thompson}}]{Wang20}%
	\BibitemOpen
	\bibfield  {author} {\bibinfo {author} {\bibfnamefont {J.}~\bibnamefont
			{Wang}}, \bibinfo {author} {\bibfnamefont {F.}~\bibnamefont {Sciarrino}},
		\bibinfo {author} {\bibfnamefont {A.}~\bibnamefont {Laing}},\ and\ \bibinfo
		{author} {\bibfnamefont {M.~G.}\ \bibnamefont {Thompson}},\ }\bibfield
	{title} {\emph {\bibinfo {title} {{\color{blue}Integrated photonic quantum
					technologies}}},\ }\href@noop {} {\bibfield  {journal} {\bibinfo  {journal}
			{Nature Photonics}\ }\textbf {\bibinfo {volume} {14}},\ \bibinfo {pages}
		{273} (\bibinfo {year} {2020})}\BibitemShut {NoStop}%
	\bibitem [{\citenamefont {Bromberg}\ \emph {et~al.}(2009)\citenamefont
		{Bromberg}, \citenamefont {Lahini}, \citenamefont {Morandotti},\ and\
		\citenamefont {Silberberg}}]{Bromberg09}%
	\BibitemOpen
	\bibfield  {author} {\bibinfo {author} {\bibfnamefont {Y.}~\bibnamefont
			{Bromberg}}, \bibinfo {author} {\bibfnamefont {Y.}~\bibnamefont {Lahini}},
		\bibinfo {author} {\bibfnamefont {R.}~\bibnamefont {Morandotti}},\ and\
		\bibinfo {author} {\bibfnamefont {Y.}~\bibnamefont {Silberberg}},\ }\bibfield
	{title} {\emph {\bibinfo {title} {{\color{blue}Quantum and classical
					correlations in waveguide lattices}}},\ }\href@noop {} {\bibfield  {journal}
		{\bibinfo  {journal} {Phys. Rev. Lett.}\ }\textbf {\bibinfo {volume} {102}},\
		\bibinfo {pages} {253904} (\bibinfo {year} {2009})}\BibitemShut {NoStop}%
	\bibitem [{\citenamefont {Peruzzo}\ \emph {et~al.}(2010)\citenamefont
		{Peruzzo}, \citenamefont {Lobino}, \citenamefont {Matthews}, \citenamefont
		{Matsuda}, \citenamefont {Politi}, \citenamefont {Poulios}, \citenamefont
		{Zhou}, \citenamefont {Lahini}, \citenamefont {Ismail}, \citenamefont
		{W{\"o}rhoff} \emph {et~al.}}]{Peruzzo10}%
	\BibitemOpen
	\bibfield  {author} {\bibinfo {author} {\bibfnamefont {A.}~\bibnamefont
			{Peruzzo}}, \bibinfo {author} {\bibfnamefont {M.}~\bibnamefont {Lobino}},
		\bibinfo {author} {\bibfnamefont {J.~C.}\ \bibnamefont {Matthews}}, \bibinfo
		{author} {\bibfnamefont {N.}~\bibnamefont {Matsuda}}, \bibinfo {author}
		{\bibfnamefont {A.}~\bibnamefont {Politi}}, \bibinfo {author} {\bibfnamefont
			{K.}~\bibnamefont {Poulios}}, \bibinfo {author} {\bibfnamefont {X.-Q.}\
			\bibnamefont {Zhou}}, \bibinfo {author} {\bibfnamefont {Y.}~\bibnamefont
			{Lahini}}, \bibinfo {author} {\bibfnamefont {N.}~\bibnamefont {Ismail}},
		\bibinfo {author} {\bibfnamefont {K.}~\bibnamefont {W{\"o}rhoff}}, \emph
		{et~al.},\ }\bibfield  {title} {\emph {\bibinfo {title} {{\color{blue}Quantum
					walks of correlated photons}}},\ }\href@noop {} {\bibfield  {journal}
		{\bibinfo  {journal} {Science}\ }\textbf {\bibinfo {volume} {329}},\ \bibinfo
		{pages} {1500} (\bibinfo {year} {2010})}\BibitemShut {NoStop}%
	\bibitem [{\citenamefont {Solntsev}\ \emph {et~al.}(2014)\citenamefont
		{Solntsev}, \citenamefont {Setzpfandt}, \citenamefont {Clark}, \citenamefont
		{Wu}, \citenamefont {Collins}, \citenamefont {Xiong}, \citenamefont
		{Schreiber}, \citenamefont {Katzschmann}, \citenamefont {Eilenberger},
		\citenamefont {Schiek}, \citenamefont {Sohler}, \citenamefont {Mitchell},
		\citenamefont {Silberhorn}, \citenamefont {Eggleton}, \citenamefont
		{Pertsch}, \citenamefont {Sukhorukov}, \citenamefont {Neshev},\ and\
		\citenamefont {Kivshar}}]{Solntsev14}%
	\BibitemOpen
	\bibfield  {author} {\bibinfo {author} {\bibfnamefont {A.~S.}\ \bibnamefont
			{Solntsev}}, \bibinfo {author} {\bibfnamefont {F.}~\bibnamefont
			{Setzpfandt}}, \bibinfo {author} {\bibfnamefont {A.~S.}\ \bibnamefont
			{Clark}}, \bibinfo {author} {\bibfnamefont {C.~W.}\ \bibnamefont {Wu}},
		\bibinfo {author} {\bibfnamefont {M.~J.}\ \bibnamefont {Collins}}, \bibinfo
		{author} {\bibfnamefont {C.}~\bibnamefont {Xiong}}, \bibinfo {author}
		{\bibfnamefont {A.}~\bibnamefont {Schreiber}}, \bibinfo {author}
		{\bibfnamefont {F.}~\bibnamefont {Katzschmann}}, \bibinfo {author}
		{\bibfnamefont {F.}~\bibnamefont {Eilenberger}}, \bibinfo {author}
		{\bibfnamefont {R.}~\bibnamefont {Schiek}}, \bibinfo {author} {\bibfnamefont
			{W.}~\bibnamefont {Sohler}}, \bibinfo {author} {\bibfnamefont
			{A.}~\bibnamefont {Mitchell}}, \bibinfo {author} {\bibfnamefont
			{C.}~\bibnamefont {Silberhorn}}, \bibinfo {author} {\bibfnamefont {B.~J.}\
			\bibnamefont {Eggleton}}, \bibinfo {author} {\bibfnamefont {T.}~\bibnamefont
			{Pertsch}}, \bibinfo {author} {\bibfnamefont {A.~A.}\ \bibnamefont
			{Sukhorukov}}, \bibinfo {author} {\bibfnamefont {D.~N.}\ \bibnamefont
			{Neshev}},\ and\ \bibinfo {author} {\bibfnamefont {Y.~S.}\ \bibnamefont
			{Kivshar}},\ }\bibfield  {title} {\emph {\bibinfo {title}
			{{\color{blue}Generation of nonclassical biphoton states through cascaded
					quantum walks on a nonlinear chip}}},\ }\href@noop {} {\bibfield  {journal}
		{\bibinfo  {journal} {Phys. Rev. X}\ }\textbf {\bibinfo {volume} {4}},\
		\bibinfo {pages} {031007} (\bibinfo {year} {2014})}\BibitemShut {NoStop}%
	\bibitem [{\citenamefont {Chapman}\ \emph {et~al.}(2016)\citenamefont
		{Chapman}, \citenamefont {Santandrea}, \citenamefont {Huang}, \citenamefont
		{Corrielli}, \citenamefont {Crespi}, \citenamefont {Yung}, \citenamefont
		{Osellame},\ and\ \citenamefont {Peruzzo}}]{Chapman16}%
	\BibitemOpen
	\bibfield  {author} {\bibinfo {author} {\bibfnamefont {R.~J.}\ \bibnamefont
			{Chapman}}, \bibinfo {author} {\bibfnamefont {M.}~\bibnamefont {Santandrea}},
		\bibinfo {author} {\bibfnamefont {Z.}~\bibnamefont {Huang}}, \bibinfo
		{author} {\bibfnamefont {G.}~\bibnamefont {Corrielli}}, \bibinfo {author}
		{\bibfnamefont {A.}~\bibnamefont {Crespi}}, \bibinfo {author} {\bibfnamefont
			{M.-H.}\ \bibnamefont {Yung}}, \bibinfo {author} {\bibfnamefont
			{R.}~\bibnamefont {Osellame}},\ and\ \bibinfo {author} {\bibfnamefont
			{A.}~\bibnamefont {Peruzzo}},\ }\bibfield  {title} {\emph {\bibinfo {title}
			{{\color{blue}Experimental perfect state transfer of an entangled photonic
					qubit}}},\ }\href@noop {} {\bibfield  {journal} {\bibinfo  {journal} {Nature
				Comm.}\ }\textbf {\bibinfo {volume} {7}},\ \bibinfo {pages} {11339} (\bibinfo
		{year} {2016})}\BibitemShut {NoStop}%
	\bibitem [{\citenamefont {Rechtsman}\ \emph {et~al.}(2016)\citenamefont
		{Rechtsman}, \citenamefont {Lumer}, \citenamefont {Plotnik}, \citenamefont
		{Perez-Leija}, \citenamefont {Szameit},\ and\ \citenamefont
		{Segev}}]{Rechtsman16}%
	\BibitemOpen
	\bibfield  {author} {\bibinfo {author} {\bibfnamefont {M.~C.}\ \bibnamefont
			{Rechtsman}}, \bibinfo {author} {\bibfnamefont {Y.}~\bibnamefont {Lumer}},
		\bibinfo {author} {\bibfnamefont {Y.}~\bibnamefont {Plotnik}}, \bibinfo
		{author} {\bibfnamefont {A.}~\bibnamefont {Perez-Leija}}, \bibinfo {author}
		{\bibfnamefont {A.}~\bibnamefont {Szameit}},\ and\ \bibinfo {author}
		{\bibfnamefont {M.}~\bibnamefont {Segev}},\ }\bibfield  {title} {\emph
		{\bibinfo {title} {{\color{blue}Topological protection of photonic path
					entanglement}}},\ }\href@noop {} {\bibfield  {journal} {\bibinfo  {journal}
			{Optica}\ }\textbf {\bibinfo {volume} {3}},\ \bibinfo {pages} {925} (\bibinfo
		{year} {2016})}\BibitemShut {NoStop}%
	\bibitem [{\citenamefont {Gr{\"a}fe}\ \emph {et~al.}(2016)\citenamefont
		{Gr{\"a}fe}, \citenamefont {Heilmann}, \citenamefont {Lebugle}, \citenamefont
		{Guzman-Silva}, \citenamefont {Perez-Leija},\ and\ \citenamefont
		{Szameit}}]{Grafe16}%
	\BibitemOpen
	\bibfield  {author} {\bibinfo {author} {\bibfnamefont {M.}~\bibnamefont
			{Gr{\"a}fe}}, \bibinfo {author} {\bibfnamefont {R.}~\bibnamefont {Heilmann}},
		\bibinfo {author} {\bibfnamefont {M.}~\bibnamefont {Lebugle}}, \bibinfo
		{author} {\bibfnamefont {D.}~\bibnamefont {Guzman-Silva}}, \bibinfo {author}
		{\bibfnamefont {A.}~\bibnamefont {Perez-Leija}},\ and\ \bibinfo {author}
		{\bibfnamefont {A.}~\bibnamefont {Szameit}},\ }\bibfield  {title} {\emph
		{\bibinfo {title} {{\color{blue}Integrated photonic quantum walks}}},\
	}\href@noop {} {\bibfield  {journal} {\bibinfo  {journal} {Journal of
				Optics}\ }\textbf {\bibinfo {volume} {18}},\ \bibinfo {pages} {103002}
		(\bibinfo {year} {2016})}\BibitemShut {NoStop}%
	\bibitem [{\citenamefont {Tambasco}\ \emph {et~al.}(2018)\citenamefont
		{Tambasco}, \citenamefont {Corrielli}, \citenamefont {Chapman}, \citenamefont
		{Crespi}, \citenamefont {Zilberberg}, \citenamefont {Osellame},\ and\
		\citenamefont {Peruzzo}}]{Tambasco18}%
	\BibitemOpen
	\bibfield  {author} {\bibinfo {author} {\bibfnamefont {J.-L.}\ \bibnamefont
			{Tambasco}}, \bibinfo {author} {\bibfnamefont {G.}~\bibnamefont {Corrielli}},
		\bibinfo {author} {\bibfnamefont {R.~J.}\ \bibnamefont {Chapman}}, \bibinfo
		{author} {\bibfnamefont {A.}~\bibnamefont {Crespi}}, \bibinfo {author}
		{\bibfnamefont {O.}~\bibnamefont {Zilberberg}}, \bibinfo {author}
		{\bibfnamefont {R.}~\bibnamefont {Osellame}},\ and\ \bibinfo {author}
		{\bibfnamefont {A.}~\bibnamefont {Peruzzo}},\ }\bibfield  {title} {\emph
		{\bibinfo {title} {{\color{blue}Quantum interference of topological states of
					light}}},\ }\href@noop {} {\bibfield  {journal} {\bibinfo  {journal} {Science
				Advances}\ }\textbf {\bibinfo {volume} {4}},\ \bibinfo {pages} {eaat3187}
		(\bibinfo {year} {2018})}\BibitemShut {NoStop}%
	\bibitem [{\citenamefont {Blanco-Redondo}\ \emph {et~al.}(2018)\citenamefont
		{Blanco-Redondo}, \citenamefont {Bell}, \citenamefont {Oren}, \citenamefont
		{Eggleton},\ and\ \citenamefont {Segev}}]{BlancoRedondo18}%
	\BibitemOpen
	\bibfield  {author} {\bibinfo {author} {\bibfnamefont {A.}~\bibnamefont
			{Blanco-Redondo}}, \bibinfo {author} {\bibfnamefont {B.}~\bibnamefont
			{Bell}}, \bibinfo {author} {\bibfnamefont {D.}~\bibnamefont {Oren}}, \bibinfo
		{author} {\bibfnamefont {B.~J.}\ \bibnamefont {Eggleton}},\ and\ \bibinfo
		{author} {\bibfnamefont {M.}~\bibnamefont {Segev}},\ }\bibfield  {title}
	{\emph {\bibinfo {title} {{\color{blue}Topological protection of biphoton
					states}}},\ }\href@noop {} {\bibfield  {journal} {\bibinfo  {journal}
			{Science}\ }\textbf {\bibinfo {volume} {362}},\ \bibinfo {pages} {568}
		(\bibinfo {year} {2018})}\BibitemShut {NoStop}%
	\bibitem [{\citenamefont {Lahini}\ \emph {et~al.}(2018)\citenamefont {Lahini},
		\citenamefont {Steinbrecher}, \citenamefont {Bookatz},\ and\ \citenamefont
		{Englund}}]{Lahini18}%
	\BibitemOpen
	\bibfield  {author} {\bibinfo {author} {\bibfnamefont {Y.}~\bibnamefont
			{Lahini}}, \bibinfo {author} {\bibfnamefont {G.~R.}\ \bibnamefont
			{Steinbrecher}}, \bibinfo {author} {\bibfnamefont {A.~D.}\ \bibnamefont
			{Bookatz}},\ and\ \bibinfo {author} {\bibfnamefont {D.}~\bibnamefont
			{Englund}},\ }\bibfield  {title} {\emph {\bibinfo {title}
			{{\color{blue}Quantum logic using correlated one-dimensional quantum
					walks}}},\ }\href@noop {} {\bibfield  {journal} {\bibinfo  {journal} {npj
				Quantum Information}\ }\textbf {\bibinfo {volume} {4}},\ \bibinfo {pages} {2}
		(\bibinfo {year} {2018})}\BibitemShut {NoStop}%
	\bibitem [{\citenamefont {Barral}\ \emph {et~al.}(2020)\citenamefont {Barral},
		\citenamefont {Walschaers}, \citenamefont {Bencheikh}, \citenamefont
		{Parigi}, \citenamefont {Levenson}, \citenamefont {Treps},\ and\
		\citenamefont {Belabas}}]{Barral20}%
	\BibitemOpen
	\bibfield  {author} {\bibinfo {author} {\bibfnamefont {D.}~\bibnamefont
			{Barral}}, \bibinfo {author} {\bibfnamefont {M.}~\bibnamefont {Walschaers}},
		\bibinfo {author} {\bibfnamefont {K.}~\bibnamefont {Bencheikh}}, \bibinfo
		{author} {\bibfnamefont {V.}~\bibnamefont {Parigi}}, \bibinfo {author}
		{\bibfnamefont {J.~A.}\ \bibnamefont {Levenson}}, \bibinfo {author}
		{\bibfnamefont {N.}~\bibnamefont {Treps}},\ and\ \bibinfo {author}
		{\bibfnamefont {N.}~\bibnamefont {Belabas}},\ }\bibfield  {title} {\emph
		{\bibinfo {title} {{\color{blue}Quantum state engineering in arrays of
					nonlinear waveguides}}},\ }\href@noop {} {\bibfield  {journal} {\bibinfo
			{journal} {Phys. Rev. A}\ }\textbf {\bibinfo {volume} {102}},\ \bibinfo
		{pages} {043706} (\bibinfo {year} {2020})}\BibitemShut {NoStop}%
	\bibitem [{\citenamefont {Barral}\ \emph {et~al.}(2021)\citenamefont {Barral},
		\citenamefont {Bencheikh}, \citenamefont {Levenson},\ and\ \citenamefont
		{Belabas}}]{Barral21}%
	\BibitemOpen
	\bibfield  {author} {\bibinfo {author} {\bibfnamefont {D.}~\bibnamefont
			{Barral}}, \bibinfo {author} {\bibfnamefont {K.}~\bibnamefont {Bencheikh}},
		\bibinfo {author} {\bibfnamefont {J.~A.}\ \bibnamefont {Levenson}},\ and\
		\bibinfo {author} {\bibfnamefont {N.}~\bibnamefont {Belabas}},\ }\bibfield
	{title} {\emph {\bibinfo {title} {{\color{blue}Scalable multimode
					entanglement based on efficient squeezing of propagation eigenmodes}}},\
	}\href@noop {} {\bibfield  {journal} {\bibinfo  {journal} {Phys. Rev. Res.}\
		}\textbf {\bibinfo {volume} {3}},\ \bibinfo {pages} {013068} (\bibinfo {year}
		{2021})}\BibitemShut {NoStop}%
	\bibitem [{\citenamefont {Hoch}\ \emph {et~al.}(2022)\citenamefont {Hoch},
		\citenamefont {Piacentini}, \citenamefont {Giordani}, \citenamefont {Tian},
		\citenamefont {Iuliano}, \citenamefont {Esposito}, \citenamefont {Camillini},
		\citenamefont {Carvacho}, \citenamefont {Ceccarelli}, \citenamefont
		{Spagnolo} \emph {et~al.}}]{Hoch22}%
	\BibitemOpen
	\bibfield  {author} {\bibinfo {author} {\bibfnamefont {F.}~\bibnamefont
			{Hoch}}, \bibinfo {author} {\bibfnamefont {S.}~\bibnamefont {Piacentini}},
		\bibinfo {author} {\bibfnamefont {T.}~\bibnamefont {Giordani}}, \bibinfo
		{author} {\bibfnamefont {Z.-N.}\ \bibnamefont {Tian}}, \bibinfo {author}
		{\bibfnamefont {M.}~\bibnamefont {Iuliano}}, \bibinfo {author} {\bibfnamefont
			{C.}~\bibnamefont {Esposito}}, \bibinfo {author} {\bibfnamefont
			{A.}~\bibnamefont {Camillini}}, \bibinfo {author} {\bibfnamefont
			{G.}~\bibnamefont {Carvacho}}, \bibinfo {author} {\bibfnamefont
			{F.}~\bibnamefont {Ceccarelli}}, \bibinfo {author} {\bibfnamefont
			{N.}~\bibnamefont {Spagnolo}}, \emph {et~al.},\ }\bibfield  {title} {\emph
		{\bibinfo {title} {{\color{blue}Reconfigurable continuously-coupled {3D}
					photonic circuit for boson sampling experiments}}},\ }\href@noop {}
	{\bibfield  {journal} {\bibinfo  {journal} {npj Quantum Information}\
		}\textbf {\bibinfo {volume} {8}},\ \bibinfo {pages} {55} (\bibinfo {year}
		{2022})}\BibitemShut {NoStop}%
	\bibitem [{\citenamefont {Raymond}\ \emph {et~al.}(2024)\citenamefont
		{Raymond}, \citenamefont {Zecchetto}, \citenamefont {Palomo}, \citenamefont
		{Morassi}, \citenamefont {Lema\^{\i}tre}, \citenamefont {Raineri},
		\citenamefont {Amanti}, \citenamefont {Ducci},\ and\ \citenamefont
		{Baboux}}]{Raymond24}%
	\BibitemOpen
	\bibfield  {author} {\bibinfo {author} {\bibfnamefont {A.}~\bibnamefont
			{Raymond}}, \bibinfo {author} {\bibfnamefont {A.}~\bibnamefont {Zecchetto}},
		\bibinfo {author} {\bibfnamefont {J.}~\bibnamefont {Palomo}}, \bibinfo
		{author} {\bibfnamefont {M.}~\bibnamefont {Morassi}}, \bibinfo {author}
		{\bibfnamefont {A.}~\bibnamefont {Lema\^{\i}tre}}, \bibinfo {author}
		{\bibfnamefont {F.}~\bibnamefont {Raineri}}, \bibinfo {author} {\bibfnamefont
			{M.~I.}\ \bibnamefont {Amanti}}, \bibinfo {author} {\bibfnamefont
			{S.}~\bibnamefont {Ducci}},\ and\ \bibinfo {author} {\bibfnamefont
			{F.}~\bibnamefont {Baboux}},\ }\bibfield  {title} {\emph {\bibinfo {title}
			{{\color{blue}Tunable Generation of Spatial Entanglement in Nonlinear
					Waveguide Arrays}}},\ }\href@noop {} {\bibfield  {journal} {\bibinfo
			{journal} {Phys. Rev. Lett.}\ }\textbf {\bibinfo {volume} {133}},\ \bibinfo
		{pages} {233602} (\bibinfo {year} {2024})}\BibitemShut {NoStop}%
	\bibitem [{\citenamefont {Raymond}\ \emph {et~al.}(2025)\citenamefont
		{Raymond}, \citenamefont {Cathala}, \citenamefont {Morassi}, \citenamefont
		{Lema\^{i}tre}, \citenamefont {Raineri}, \citenamefont {Ducci},\ and\
		\citenamefont {Baboux}}]{Raymond25}%
	\BibitemOpen
	\bibfield  {author} {\bibinfo {author} {\bibfnamefont {A.}~\bibnamefont
			{Raymond}}, \bibinfo {author} {\bibfnamefont {P.}~\bibnamefont {Cathala}},
		\bibinfo {author} {\bibfnamefont {M.}~\bibnamefont {Morassi}}, \bibinfo
		{author} {\bibfnamefont {A.}~\bibnamefont {Lema\^{i}tre}}, \bibinfo {author}
		{\bibfnamefont {F.}~\bibnamefont {Raineri}}, \bibinfo {author} {\bibfnamefont
			{S.}~\bibnamefont {Ducci}},\ and\ \bibinfo {author} {\bibfnamefont
			{F.}~\bibnamefont {Baboux}},\ }\bibfield  {title} {\emph {\bibinfo {title}
			{{\color{blue}Tailoring quantum walks in integrated photonic lattices}}},\
	}\href@noop {} {\bibfield  {journal} {\bibinfo  {journal} {Opt. Express}\
		}\textbf {\bibinfo {volume} {33}},\ \bibinfo {pages} {45869} (\bibinfo {year}
		{2025})}\BibitemShut {NoStop}%
	\bibitem [{\citenamefont {Chapman}\ \emph {et~al.}(2023)\citenamefont
		{Chapman}, \citenamefont {Häusler}, \citenamefont {Finco}, \citenamefont
		{Kaufmann},\ and\ \citenamefont {Grange}}]{Chapman24}%
	\BibitemOpen
	\bibfield  {author} {\bibinfo {author} {\bibfnamefont {R.~J.}\ \bibnamefont
			{Chapman}}, \bibinfo {author} {\bibfnamefont {S.}~\bibnamefont {Häusler}},
		\bibinfo {author} {\bibfnamefont {G.}~\bibnamefont {Finco}}, \bibinfo
		{author} {\bibfnamefont {F.}~\bibnamefont {Kaufmann}},\ and\ \bibinfo
		{author} {\bibfnamefont {R.}~\bibnamefont {Grange}},\ }\bibfield  {title}
	{\emph {\bibinfo {title} {{\color{blue}Quantum logical controlled-NOT gate in
					a lithium niobate-on-insulator photonic quantum walk}}},\ }\href@noop {}
	{\bibfield  {journal} {\bibinfo  {journal} {Quantum Science and Technology}\
		}\textbf {\bibinfo {volume} {9}},\ \bibinfo {pages} {015016} (\bibinfo {year}
		{2023})}\BibitemShut {NoStop}%
	\bibitem [{\citenamefont {Yang}\ \emph {et~al.}(2025)\citenamefont {Yang},
		\citenamefont {Chapman}, \citenamefont {Youssry}, \citenamefont {Haylock},
		\citenamefont {Lenzini}, \citenamefont {Lobino},\ and\ \citenamefont
		{Peruzzo}}]{Yang25}%
	\BibitemOpen
	\bibfield  {author} {\bibinfo {author} {\bibfnamefont {Y.}~\bibnamefont
			{Yang}}, \bibinfo {author} {\bibfnamefont {R.~J.}\ \bibnamefont {Chapman}},
		\bibinfo {author} {\bibfnamefont {A.}~\bibnamefont {Youssry}}, \bibinfo
		{author} {\bibfnamefont {B.}~\bibnamefont {Haylock}}, \bibinfo {author}
		{\bibfnamefont {F.}~\bibnamefont {Lenzini}}, \bibinfo {author} {\bibfnamefont
			{M.}~\bibnamefont {Lobino}},\ and\ \bibinfo {author} {\bibfnamefont
			{A.}~\bibnamefont {Peruzzo}},\ }\bibfield  {title} {\emph {\bibinfo {title}
			{{\color{blue}Programmable quantum circuits in a large-scale photonic
					waveguide array}}},\ }\href@noop {} {\bibfield  {journal} {\bibinfo
			{journal} {npj Quantum Information}\ }\textbf {\bibinfo {volume} {11}},\
		\bibinfo {pages} {19} (\bibinfo {year} {2025})}\BibitemShut {NoStop}%
	\bibitem [{\citenamefont {Perets}\ \emph {et~al.}(2008)\citenamefont {Perets},
		\citenamefont {Lahini}, \citenamefont {Pozzi}, \citenamefont {Sorel},
		\citenamefont {Morandotti},\ and\ \citenamefont {Silberberg}}]{Perets08}%
	\BibitemOpen
	\bibfield  {author} {\bibinfo {author} {\bibfnamefont {H.~B.}\ \bibnamefont
			{Perets}}, \bibinfo {author} {\bibfnamefont {Y.}~\bibnamefont {Lahini}},
		\bibinfo {author} {\bibfnamefont {F.}~\bibnamefont {Pozzi}}, \bibinfo
		{author} {\bibfnamefont {M.}~\bibnamefont {Sorel}}, \bibinfo {author}
		{\bibfnamefont {R.}~\bibnamefont {Morandotti}},\ and\ \bibinfo {author}
		{\bibfnamefont {Y.}~\bibnamefont {Silberberg}},\ }\bibfield  {title} {\emph
		{\bibinfo {title} {{\color{blue}Realization of quantum walks with negligible
					decoherence in waveguide lattices}}},\ }\href@noop {} {\bibfield  {journal}
		{\bibinfo  {journal} {Phys. Rev. Lett.}\ }\textbf {\bibinfo {volume} {100}},\
		\bibinfo {pages} {170506} (\bibinfo {year} {2008})}\BibitemShut {NoStop}%
	\bibitem [{\citenamefont {Di~Giuseppe}\ \emph {et~al.}(2013)\citenamefont
		{Di~Giuseppe}, \citenamefont {Martin}, \citenamefont {Perez-Leija},
		\citenamefont {Keil}, \citenamefont {Dreisow}, \citenamefont {Nolte},
		\citenamefont {Szameit}, \citenamefont {Abouraddy}, \citenamefont
		{Christodoulides},\ and\ \citenamefont {Saleh}}]{DiGiuseppe13}%
	\BibitemOpen
	\bibfield  {author} {\bibinfo {author} {\bibfnamefont {G.}~\bibnamefont
			{Di~Giuseppe}}, \bibinfo {author} {\bibfnamefont {L.}~\bibnamefont {Martin}},
		\bibinfo {author} {\bibfnamefont {A.}~\bibnamefont {Perez-Leija}}, \bibinfo
		{author} {\bibfnamefont {R.}~\bibnamefont {Keil}}, \bibinfo {author}
		{\bibfnamefont {F.}~\bibnamefont {Dreisow}}, \bibinfo {author} {\bibfnamefont
			{S.}~\bibnamefont {Nolte}}, \bibinfo {author} {\bibfnamefont
			{A.}~\bibnamefont {Szameit}}, \bibinfo {author} {\bibfnamefont {A.~F.}\
			\bibnamefont {Abouraddy}}, \bibinfo {author} {\bibfnamefont {D.~N.}\
			\bibnamefont {Christodoulides}},\ and\ \bibinfo {author} {\bibfnamefont
			{B.~E.~A.}\ \bibnamefont {Saleh}},\ }\bibfield  {title} {\emph {\bibinfo
			{title} {{\color{blue}{Einstein-Podolsky-Rosen} spatial entanglement in
					ordered and {Anderson} photonic lattices}}},\ }\href@noop {} {\bibfield
		{journal} {\bibinfo  {journal} {Phys. Rev. Lett.}\ }\textbf {\bibinfo
			{volume} {110}},\ \bibinfo {pages} {150503} (\bibinfo {year}
		{2013})}\BibitemShut {NoStop}%
	\bibitem [{\citenamefont {Lebugle}\ \emph {et~al.}(2015)\citenamefont
		{Lebugle}, \citenamefont {Gr{\"{a}}fe}, \citenamefont {Heilmann},
		\citenamefont {Perez-Leija}, \citenamefont {Nolte},\ and\ \citenamefont
		{Szameit}}]{Lebugle15}%
	\BibitemOpen
	\bibfield  {author} {\bibinfo {author} {\bibfnamefont {M.}~\bibnamefont
			{Lebugle}}, \bibinfo {author} {\bibfnamefont {M.}~\bibnamefont
			{Gr{\"{a}}fe}}, \bibinfo {author} {\bibfnamefont {R.}~\bibnamefont
			{Heilmann}}, \bibinfo {author} {\bibfnamefont {A.}~\bibnamefont
			{Perez-Leija}}, \bibinfo {author} {\bibfnamefont {S.}~\bibnamefont {Nolte}},\
		and\ \bibinfo {author} {\bibfnamefont {A.}~\bibnamefont {Szameit}},\
	}\bibfield  {title} {\emph {\bibinfo {title} {{\color{blue}{Experimental
						observation of N00N state Bloch oscillations}}}},\ }\href@noop {} {\bibfield
		{journal} {\bibinfo  {journal} {Nature Communications}\ }\textbf {\bibinfo
			{volume} {6}},\ \bibinfo {pages} {8273} (\bibinfo {year} {2015})}\BibitemShut
	{NoStop}%
	\bibitem [{\citenamefont {Caruso}\ \emph {et~al.}(2016)\citenamefont {Caruso},
		\citenamefont {Crespi}, \citenamefont {Ciriolo}, \citenamefont {Sciarrino},\
		and\ \citenamefont {Osellame}}]{Caruso16}%
	\BibitemOpen
	\bibfield  {author} {\bibinfo {author} {\bibfnamefont {F.}~\bibnamefont
			{Caruso}}, \bibinfo {author} {\bibfnamefont {A.}~\bibnamefont {Crespi}},
		\bibinfo {author} {\bibfnamefont {A.~G.}\ \bibnamefont {Ciriolo}}, \bibinfo
		{author} {\bibfnamefont {F.}~\bibnamefont {Sciarrino}},\ and\ \bibinfo
		{author} {\bibfnamefont {R.}~\bibnamefont {Osellame}},\ }\bibfield  {title}
	{\emph {\bibinfo {title} {{\color{blue}Fast escape of a quantum walker from
					an integrated photonic maze}}},\ }\href@noop {} {\bibfield  {journal}
		{\bibinfo  {journal} {Nature Comm.}\ }\textbf {\bibinfo {volume} {7}},\
		\bibinfo {pages} {11682} (\bibinfo {year} {2016})}\BibitemShut {NoStop}%
	\bibitem [{\citenamefont {Biggerstaff}\ \emph {et~al.}(2016)\citenamefont
		{Biggerstaff}, \citenamefont {Heilmann}, \citenamefont {Zecevik},
		\citenamefont {Gr{\"a}fe}, \citenamefont {Broome}, \citenamefont {Fedrizzi},
		\citenamefont {Nolte}, \citenamefont {Szameit}, \citenamefont {White},\ and\
		\citenamefont {Kassal}}]{Biggerstaff16}%
	\BibitemOpen
	\bibfield  {author} {\bibinfo {author} {\bibfnamefont {D.~N.}\ \bibnamefont
			{Biggerstaff}}, \bibinfo {author} {\bibfnamefont {R.}~\bibnamefont
			{Heilmann}}, \bibinfo {author} {\bibfnamefont {A.~A.}\ \bibnamefont
			{Zecevik}}, \bibinfo {author} {\bibfnamefont {M.}~\bibnamefont {Gr{\"a}fe}},
		\bibinfo {author} {\bibfnamefont {M.~A.}\ \bibnamefont {Broome}}, \bibinfo
		{author} {\bibfnamefont {A.}~\bibnamefont {Fedrizzi}}, \bibinfo {author}
		{\bibfnamefont {S.}~\bibnamefont {Nolte}}, \bibinfo {author} {\bibfnamefont
			{A.}~\bibnamefont {Szameit}}, \bibinfo {author} {\bibfnamefont {A.~G.}\
			\bibnamefont {White}},\ and\ \bibinfo {author} {\bibfnamefont
			{I.}~\bibnamefont {Kassal}},\ }\bibfield  {title} {\emph {\bibinfo {title}
			{{\color{blue}Enhancing coherent transport in a photonic network using
					controllable decoherence}}},\ }\href@noop {} {\bibfield  {journal} {\bibinfo
			{journal} {Nature Comm.}\ }\textbf {\bibinfo {volume} {7}},\ \bibinfo {pages}
		{11282} (\bibinfo {year} {2016})}\BibitemShut {NoStop}%
	\bibitem [{\citenamefont {Ozawa}\ \emph {et~al.}(2019)\citenamefont {Ozawa},
		\citenamefont {Price}, \citenamefont {Amo}, \citenamefont {Goldman},
		\citenamefont {Hafezi}, \citenamefont {Lu}, \citenamefont {Rechtsman},
		\citenamefont {Schuster}, \citenamefont {Simon}, \citenamefont {Zilberberg}
		\emph {et~al.}}]{Ozawa19}%
	\BibitemOpen
	\bibfield  {author} {\bibinfo {author} {\bibfnamefont {T.}~\bibnamefont
			{Ozawa}}, \bibinfo {author} {\bibfnamefont {H.~M.}\ \bibnamefont {Price}},
		\bibinfo {author} {\bibfnamefont {A.}~\bibnamefont {Amo}}, \bibinfo {author}
		{\bibfnamefont {N.}~\bibnamefont {Goldman}}, \bibinfo {author} {\bibfnamefont
			{M.}~\bibnamefont {Hafezi}}, \bibinfo {author} {\bibfnamefont
			{L.}~\bibnamefont {Lu}}, \bibinfo {author} {\bibfnamefont {M.~C.}\
			\bibnamefont {Rechtsman}}, \bibinfo {author} {\bibfnamefont {D.}~\bibnamefont
			{Schuster}}, \bibinfo {author} {\bibfnamefont {J.}~\bibnamefont {Simon}},
		\bibinfo {author} {\bibfnamefont {O.}~\bibnamefont {Zilberberg}}, \emph
		{et~al.},\ }\bibfield  {title} {\emph {\bibinfo {title}
			{{\color{blue}Topological photonics}}},\ }\href@noop {} {\bibfield  {journal}
		{\bibinfo  {journal} {Rev. Mod. Phys.}\ }\textbf {\bibinfo {volume} {91}},\
		\bibinfo {pages} {015006} (\bibinfo {year} {2019})}\BibitemShut {NoStop}%
	\bibitem [{\citenamefont {Gao}\ \emph {et~al.}(2024)\citenamefont {Gao},
		\citenamefont {Xu}, \citenamefont {Yang}, \citenamefont {Zwiller},\ and\
		\citenamefont {Elshaari}}]{Gao24_review}%
	\BibitemOpen
	\bibfield  {author} {\bibinfo {author} {\bibfnamefont {J.}~\bibnamefont
			{Gao}}, \bibinfo {author} {\bibfnamefont {Z.-S.}\ \bibnamefont {Xu}},
		\bibinfo {author} {\bibfnamefont {Z.}~\bibnamefont {Yang}}, \bibinfo {author}
		{\bibfnamefont {V.}~\bibnamefont {Zwiller}},\ and\ \bibinfo {author}
		{\bibfnamefont {A.~W.}\ \bibnamefont {Elshaari}},\ }\bibfield  {title} {\emph
		{\bibinfo {title} {{\color{blue}Quantum topological photonics with special
					focus on waveguide systems}}},\ }\href@noop {} {\bibfield  {journal}
		{\bibinfo  {journal} {npj Nanophotonics}\ }\textbf {\bibinfo {volume} {1}},\
		\bibinfo {pages} {34} (\bibinfo {year} {2024})}\BibitemShut {NoStop}%
	\bibitem [{\citenamefont {Hashemi}\ \emph {et~al.}(2025)\citenamefont
		{Hashemi}, \citenamefont {Zakeri}, \citenamefont {Jung},\ and\ \citenamefont
		{Blanco-Redondo}}]{Hashemi25}%
	\BibitemOpen
	\bibfield  {author} {\bibinfo {author} {\bibfnamefont {A.}~\bibnamefont
			{Hashemi}}, \bibinfo {author} {\bibfnamefont {M.~J.}\ \bibnamefont {Zakeri}},
		\bibinfo {author} {\bibfnamefont {P.~S.}\ \bibnamefont {Jung}},\ and\
		\bibinfo {author} {\bibfnamefont {A.}~\bibnamefont {Blanco-Redondo}},\
	}\bibfield  {title} {\emph {\bibinfo {title} {{\color{blue}Topological
					quantum photonics}}},\ }\href@noop {} {\bibfield  {journal} {\bibinfo
			{journal} {APL Photonics}\ }\textbf {\bibinfo {volume} {10}} (\bibinfo {year}
		{2025})}\BibitemShut {NoStop}%
	\bibitem [{\citenamefont {Wang}\ \emph
		{et~al.}(2019{\natexlab{a}})\citenamefont {Wang}, \citenamefont {Pang},
		\citenamefont {Lu}, \citenamefont {Gao}, \citenamefont {Chang}, \citenamefont
		{Qiao}, \citenamefont {Jiao}, \citenamefont {Tang},\ and\ \citenamefont
		{Jin}}]{Wang19}%
	\BibitemOpen
	\bibfield  {author} {\bibinfo {author} {\bibfnamefont {Y.}~\bibnamefont
			{Wang}}, \bibinfo {author} {\bibfnamefont {X.-L.}\ \bibnamefont {Pang}},
		\bibinfo {author} {\bibfnamefont {Y.-H.}\ \bibnamefont {Lu}}, \bibinfo
		{author} {\bibfnamefont {J.}~\bibnamefont {Gao}}, \bibinfo {author}
		{\bibfnamefont {Y.-J.}\ \bibnamefont {Chang}}, \bibinfo {author}
		{\bibfnamefont {L.-F.}\ \bibnamefont {Qiao}}, \bibinfo {author}
		{\bibfnamefont {Z.-Q.}\ \bibnamefont {Jiao}}, \bibinfo {author}
		{\bibfnamefont {H.}~\bibnamefont {Tang}},\ and\ \bibinfo {author}
		{\bibfnamefont {X.-M.}\ \bibnamefont {Jin}},\ }\bibfield  {title} {\emph
		{\bibinfo {title} {{\color{blue}Topological protection of two-photon quantum
					correlation on a photonic chip}}},\ }\href@noop {} {\bibfield  {journal}
		{\bibinfo  {journal} {Optica}\ }\textbf {\bibinfo {volume} {6}},\ \bibinfo
		{pages} {955} (\bibinfo {year} {2019}{\natexlab{a}})}\BibitemShut {NoStop}%
	\bibitem [{\citenamefont {Wang}\ \emph {et~al.}(2022)\citenamefont {Wang},
		\citenamefont {Lu}, \citenamefont {Gao}, \citenamefont {Chang}, \citenamefont
		{Ren}, \citenamefont {Jiao}, \citenamefont {Zhang},\ and\ \citenamefont
		{Jin}}]{Wang22}%
	\BibitemOpen
	\bibfield  {author} {\bibinfo {author} {\bibfnamefont {Y.}~\bibnamefont
			{Wang}}, \bibinfo {author} {\bibfnamefont {Y.-H.}\ \bibnamefont {Lu}},
		\bibinfo {author} {\bibfnamefont {J.}~\bibnamefont {Gao}}, \bibinfo {author}
		{\bibfnamefont {Y.-J.}\ \bibnamefont {Chang}}, \bibinfo {author}
		{\bibfnamefont {R.-J.}\ \bibnamefont {Ren}}, \bibinfo {author} {\bibfnamefont
			{Z.-Q.}\ \bibnamefont {Jiao}}, \bibinfo {author} {\bibfnamefont {Z.-Y.}\
			\bibnamefont {Zhang}},\ and\ \bibinfo {author} {\bibfnamefont {X.-M.}\
			\bibnamefont {Jin}},\ }\bibfield  {title} {\emph {\bibinfo {title}
			{{\color{blue}Topologically protected polarization quantum entanglement on a
					photonic chip}}},\ }\href@noop {} {\bibfield  {journal} {\bibinfo  {journal}
			{Chip}\ }\textbf {\bibinfo {volume} {1}},\ \bibinfo {pages} {100003}
		(\bibinfo {year} {2022})}\BibitemShut {NoStop}%
	\bibitem [{\citenamefont {Leykam}\ \emph {et~al.}(2015)\citenamefont {Leykam},
		\citenamefont {Solntsev}, \citenamefont {Sukhorukov},\ and\ \citenamefont
		{Desyatnikov}}]{Leykam15}%
	\BibitemOpen
	\bibfield  {author} {\bibinfo {author} {\bibfnamefont {D.}~\bibnamefont
			{Leykam}}, \bibinfo {author} {\bibfnamefont {A.~S.}\ \bibnamefont
			{Solntsev}}, \bibinfo {author} {\bibfnamefont {A.~A.}\ \bibnamefont
			{Sukhorukov}},\ and\ \bibinfo {author} {\bibfnamefont {A.~S.}\ \bibnamefont
			{Desyatnikov}},\ }\bibfield  {title} {\emph {\bibinfo {title}
			{{\color{blue}Lattice topology and spontaneous parametric down-conversion in
					quadratic nonlinear waveguide arrays}}},\ }\href@noop {} {\bibfield
		{journal} {\bibinfo  {journal} {Phys. Rev. A}\ }\textbf {\bibinfo {volume}
			{92}},\ \bibinfo {pages} {033815} (\bibinfo {year} {2015})}\BibitemShut
	{NoStop}%
	\bibitem [{\citenamefont {Yang}\ \emph {et~al.}(2021)\citenamefont {Yang},
		\citenamefont {Zuo},\ and\ \citenamefont {Cao}}]{Yang21}%
	\BibitemOpen
	\bibfield  {author} {\bibinfo {author} {\bibfnamefont {Y.}~\bibnamefont
			{Yang}}, \bibinfo {author} {\bibfnamefont {Z.}~\bibnamefont {Zuo}},\ and\
		\bibinfo {author} {\bibfnamefont {D.}~\bibnamefont {Cao}},\ }\bibfield
	{title} {\emph {\bibinfo {title} {{\color{blue}Biphoton topology in a
					quadratic nonlinear waveguide array under the Su-Schrieffer-Heeger model}}},\
	}\href@noop {} {\bibfield  {journal} {\bibinfo  {journal} {Phys. Rev. A}\
		}\textbf {\bibinfo {volume} {104}},\ \bibinfo {pages} {043710} (\bibinfo
		{year} {2021})}\BibitemShut {NoStop}%
	\bibitem [{\citenamefont {Wang}\ \emph {et~al.}(2026)\citenamefont {Wang},
		\citenamefont {Qiu}, \citenamefont {Zhao}, \citenamefont {Jing},
		\citenamefont {Zheng}, \citenamefont {Zhao}, \citenamefont {Tao},
		\citenamefont {Liu}, \citenamefont {Sheng},\ and\ \citenamefont
		{Lu}}]{Wang26}%
	\BibitemOpen
	\bibfield  {author} {\bibinfo {author} {\bibfnamefont {D.}~\bibnamefont
			{Wang}}, \bibinfo {author} {\bibfnamefont {Q.}~\bibnamefont {Qiu}}, \bibinfo
		{author} {\bibfnamefont {Y.}~\bibnamefont {Zhao}}, \bibinfo {author}
		{\bibfnamefont {X.}~\bibnamefont {Jing}}, \bibinfo {author} {\bibfnamefont
			{X.}~\bibnamefont {Zheng}}, \bibinfo {author} {\bibfnamefont
			{S.}~\bibnamefont {Zhao}}, \bibinfo {author} {\bibfnamefont {Y.}~\bibnamefont
			{Tao}}, \bibinfo {author} {\bibfnamefont {G.}~\bibnamefont {Liu}}, \bibinfo
		{author} {\bibfnamefont {C.}~\bibnamefont {Sheng}},\ and\ \bibinfo {author}
		{\bibfnamefont {L.}~\bibnamefont {Lu}},\ }\bibfield  {title} {\emph {\bibinfo
			{title} {{\color{blue}Topological localization of biphoton states in arrays
					of nonlinear photonic waveguides}}},\ }\href@noop {} {\bibfield  {journal}
		{\bibinfo  {journal} {Phys. Rev. A}\ }\textbf {\bibinfo {volume} {113}},\
		\bibinfo {pages} {023502} (\bibinfo {year} {2026})}\BibitemShut {NoStop}%
	\bibitem [{\citenamefont {Ren}\ \emph {et~al.}(2022)\citenamefont {Ren},
		\citenamefont {Lu}, \citenamefont {Jiang}, \citenamefont {Gao}, \citenamefont
		{Zhou}, \citenamefont {Wang}, \citenamefont {Jiao}, \citenamefont {Wang},
		\citenamefont {Solntsev},\ and\ \citenamefont {Jin}}]{Ren22}%
	\BibitemOpen
	\bibfield  {author} {\bibinfo {author} {\bibfnamefont {R.-J.}\ \bibnamefont
			{Ren}}, \bibinfo {author} {\bibfnamefont {Y.-H.}\ \bibnamefont {Lu}},
		\bibinfo {author} {\bibfnamefont {Z.-K.}\ \bibnamefont {Jiang}}, \bibinfo
		{author} {\bibfnamefont {J.}~\bibnamefont {Gao}}, \bibinfo {author}
		{\bibfnamefont {W.-H.}\ \bibnamefont {Zhou}}, \bibinfo {author}
		{\bibfnamefont {Y.}~\bibnamefont {Wang}}, \bibinfo {author} {\bibfnamefont
			{Z.-Q.}\ \bibnamefont {Jiao}}, \bibinfo {author} {\bibfnamefont {X.-W.}\
			\bibnamefont {Wang}}, \bibinfo {author} {\bibfnamefont {A.~S.}\ \bibnamefont
			{Solntsev}},\ and\ \bibinfo {author} {\bibfnamefont {X.-M.}\ \bibnamefont
			{Jin}},\ }\bibfield  {title} {\emph {\bibinfo {title}
			{{\color{blue}Topologically protecting squeezed light on a photonic chip}}},\
	}\href@noop {} {\bibfield  {journal} {\bibinfo  {journal} {Photonics
				Research}\ }\textbf {\bibinfo {volume} {10}},\ \bibinfo {pages} {456}
		(\bibinfo {year} {2022})}\BibitemShut {NoStop}%
	\bibitem [{\citenamefont {Wang}\ \emph
		{et~al.}(2019{\natexlab{b}})\citenamefont {Wang}, \citenamefont {Doyle},
		\citenamefont {Bell}, \citenamefont {Collins}, \citenamefont {Magi},
		\citenamefont {Eggleton}, \citenamefont {Segev},\ and\ \citenamefont
		{Blanco-Redondo}}]{Wang19Blanco}%
	\BibitemOpen
	\bibfield  {author} {\bibinfo {author} {\bibfnamefont {M.}~\bibnamefont
			{Wang}}, \bibinfo {author} {\bibfnamefont {C.}~\bibnamefont {Doyle}},
		\bibinfo {author} {\bibfnamefont {B.}~\bibnamefont {Bell}}, \bibinfo {author}
		{\bibfnamefont {M.~J.}\ \bibnamefont {Collins}}, \bibinfo {author}
		{\bibfnamefont {E.}~\bibnamefont {Magi}}, \bibinfo {author} {\bibfnamefont
			{B.~J.}\ \bibnamefont {Eggleton}}, \bibinfo {author} {\bibfnamefont
			{M.}~\bibnamefont {Segev}},\ and\ \bibinfo {author} {\bibfnamefont
			{A.}~\bibnamefont {Blanco-Redondo}},\ }\bibfield  {title} {\emph {\bibinfo
			{title} {{\color{blue}Topologically protected entangled photonic states}}},\
	}\href@noop {} {\bibfield  {journal} {\bibinfo  {journal} {Nanophotonics}\
		}\textbf {\bibinfo {volume} {8}},\ \bibinfo {pages} {1327} (\bibinfo {year}
		{2019}{\natexlab{b}})}\BibitemShut {NoStop}%
	\bibitem [{\citenamefont {Bergamasco}\ and\ \citenamefont
		{Liscidini}(2019)}]{Bergamasco19}%
	\BibitemOpen
	\bibfield  {author} {\bibinfo {author} {\bibfnamefont {N.}~\bibnamefont
			{Bergamasco}}\ and\ \bibinfo {author} {\bibfnamefont {M.}~\bibnamefont
			{Liscidini}},\ }\bibfield  {title} {\emph {\bibinfo {title}
			{{\color{blue}Generation of photon pairs in topologically protected guided
					modes}}},\ }\href@noop {} {\bibfield  {journal} {\bibinfo  {journal} {Phys.
				Rev. A}\ }\textbf {\bibinfo {volume} {100}},\ \bibinfo {pages} {053827}
		(\bibinfo {year} {2019})}\BibitemShut {NoStop}%
	\bibitem [{\citenamefont {Mittal}\ \emph {et~al.}(2018)\citenamefont {Mittal},
		\citenamefont {Goldschmidt},\ and\ \citenamefont {Hafezi}}]{Mittal18}%
	\BibitemOpen
	\bibfield  {author} {\bibinfo {author} {\bibfnamefont {S.}~\bibnamefont
			{Mittal}}, \bibinfo {author} {\bibfnamefont {E.~A.}\ \bibnamefont
			{Goldschmidt}},\ and\ \bibinfo {author} {\bibfnamefont {M.}~\bibnamefont
			{Hafezi}},\ }\bibfield  {title} {\emph {\bibinfo {title} {{\color{blue}A
					topological source of quantum light}}},\ }\href@noop {} {\bibfield  {journal}
		{\bibinfo  {journal} {Nature}\ }\textbf {\bibinfo {volume} {561}},\ \bibinfo
		{pages} {502} (\bibinfo {year} {2018})}\BibitemShut {NoStop}%
	\bibitem [{\citenamefont {Silverstone}\ \emph {et~al.}(2014)\citenamefont
		{Silverstone}, \citenamefont {Bonneau}, \citenamefont {Ohira}, \citenamefont
		{Suzuki}, \citenamefont {Yoshida}, \citenamefont {Iizuka}, \citenamefont
		{Ezaki}, \citenamefont {Natarajan}, \citenamefont {Tanner}, \citenamefont
		{Hadfield} \emph {et~al.}}]{Silverstone14}%
	\BibitemOpen
	\bibfield  {author} {\bibinfo {author} {\bibfnamefont {J.~W.}\ \bibnamefont
			{Silverstone}}, \bibinfo {author} {\bibfnamefont {D.}~\bibnamefont
			{Bonneau}}, \bibinfo {author} {\bibfnamefont {K.}~\bibnamefont {Ohira}},
		\bibinfo {author} {\bibfnamefont {N.}~\bibnamefont {Suzuki}}, \bibinfo
		{author} {\bibfnamefont {H.}~\bibnamefont {Yoshida}}, \bibinfo {author}
		{\bibfnamefont {N.}~\bibnamefont {Iizuka}}, \bibinfo {author} {\bibfnamefont
			{M.}~\bibnamefont {Ezaki}}, \bibinfo {author} {\bibfnamefont {C.~M.}\
			\bibnamefont {Natarajan}}, \bibinfo {author} {\bibfnamefont {M.~G.}\
			\bibnamefont {Tanner}}, \bibinfo {author} {\bibfnamefont {R.~H.}\
			\bibnamefont {Hadfield}}, \emph {et~al.},\ }\bibfield  {title} {\emph
		{\bibinfo {title} {{\color{blue}On-chip quantum interference between silicon
					photon-pair sources}}},\ }\href@noop {} {\bibfield  {journal} {\bibinfo
			{journal} {Nature Photonics}\ }\textbf {\bibinfo {volume} {8}},\ \bibinfo
		{pages} {104} (\bibinfo {year} {2014})}\BibitemShut {NoStop}%
	\bibitem [{\citenamefont {Wang}\ \emph {et~al.}(2018)\citenamefont {Wang},
		\citenamefont {Paesani}, \citenamefont {Ding}, \citenamefont {Santagati},
		\citenamefont {Skrzypczyk}, \citenamefont {Salavrakos}, \citenamefont {Tura},
		\citenamefont {Augusiak}, \citenamefont {Man{\v{c}}inska}, \citenamefont
		{Bacco} \emph {et~al.}}]{Wang18}%
	\BibitemOpen
	\bibfield  {author} {\bibinfo {author} {\bibfnamefont {J.}~\bibnamefont
			{Wang}}, \bibinfo {author} {\bibfnamefont {S.}~\bibnamefont {Paesani}},
		\bibinfo {author} {\bibfnamefont {Y.}~\bibnamefont {Ding}}, \bibinfo {author}
		{\bibfnamefont {R.}~\bibnamefont {Santagati}}, \bibinfo {author}
		{\bibfnamefont {P.}~\bibnamefont {Skrzypczyk}}, \bibinfo {author}
		{\bibfnamefont {A.}~\bibnamefont {Salavrakos}}, \bibinfo {author}
		{\bibfnamefont {J.}~\bibnamefont {Tura}}, \bibinfo {author} {\bibfnamefont
			{R.}~\bibnamefont {Augusiak}}, \bibinfo {author} {\bibfnamefont
			{L.}~\bibnamefont {Man{\v{c}}inska}}, \bibinfo {author} {\bibfnamefont
			{D.}~\bibnamefont {Bacco}}, \emph {et~al.},\ }\bibfield  {title} {\emph
		{\bibinfo {title} {{\color{blue}Multidimensional quantum entanglement with
					large-scale integrated optics}}},\ }\href@noop {} {\bibfield  {journal}
		{\bibinfo  {journal} {Science}\ }\textbf {\bibinfo {volume} {360}},\ \bibinfo
		{pages} {285} (\bibinfo {year} {2018})}\BibitemShut {NoStop}%
	\bibitem [{\citenamefont {Chapman}\ \emph {et~al.}(2025)\citenamefont
		{Chapman}, \citenamefont {Kuttner}, \citenamefont {Kellner}, \citenamefont
		{Sabatti}, \citenamefont {Maeder}, \citenamefont {Finco}, \citenamefont
		{Kaufmann},\ and\ \citenamefont {Grange}}]{Chapman25}%
	\BibitemOpen
	\bibfield  {author} {\bibinfo {author} {\bibfnamefont {R.~J.}\ \bibnamefont
			{Chapman}}, \bibinfo {author} {\bibfnamefont {T.}~\bibnamefont {Kuttner}},
		\bibinfo {author} {\bibfnamefont {J.}~\bibnamefont {Kellner}}, \bibinfo
		{author} {\bibfnamefont {A.}~\bibnamefont {Sabatti}}, \bibinfo {author}
		{\bibfnamefont {A.}~\bibnamefont {Maeder}}, \bibinfo {author} {\bibfnamefont
			{G.}~\bibnamefont {Finco}}, \bibinfo {author} {\bibfnamefont
			{F.}~\bibnamefont {Kaufmann}},\ and\ \bibinfo {author} {\bibfnamefont
			{R.}~\bibnamefont {Grange}},\ }\bibfield  {title} {\emph {\bibinfo {title}
			{{\color{blue}On-chip quantum interference between independent lithium
					niobate-on-insulator photon-pair sources}}},\ }\href@noop {} {\bibfield
		{journal} {\bibinfo  {journal} {Physical Review Letters}\ }\textbf {\bibinfo
			{volume} {134}},\ \bibinfo {pages} {223602} (\bibinfo {year}
		{2025})}\BibitemShut {NoStop}%
	\bibitem [{\citenamefont {Solntsev}\ \emph {et~al.}(2012)\citenamefont
		{Solntsev}, \citenamefont {Sukhorukov}, \citenamefont {Neshev},\ and\
		\citenamefont {Kivshar}}]{Solntsev12}%
	\BibitemOpen
	\bibfield  {author} {\bibinfo {author} {\bibfnamefont {A.~S.}\ \bibnamefont
			{Solntsev}}, \bibinfo {author} {\bibfnamefont {A.~A.}\ \bibnamefont
			{Sukhorukov}}, \bibinfo {author} {\bibfnamefont {D.~N.}\ \bibnamefont
			{Neshev}},\ and\ \bibinfo {author} {\bibfnamefont {Y.~S.}\ \bibnamefont
			{Kivshar}},\ }\bibfield  {title} {\emph {\bibinfo {title}
			{{\color{blue}Spontaneous parametric down-conversion and quantum walks in
					arrays of quadratic nonlinear waveguides}}},\ }\href@noop {} {\bibfield
		{journal} {\bibinfo  {journal} {Phys. Rev. Lett.}\ }\textbf {\bibinfo
			{volume} {108}},\ \bibinfo {pages} {023601} (\bibinfo {year}
		{2012})}\BibitemShut {NoStop}%
	\bibitem [{\citenamefont {Gr{\"a}fe}\ \emph {et~al.}(2012)\citenamefont
		{Gr{\"a}fe}, \citenamefont {Solntsev}, \citenamefont {Keil}, \citenamefont
		{Sukhorukov}, \citenamefont {Heinrich}, \citenamefont {T{\"u}nnermann},
		\citenamefont {Nolte}, \citenamefont {Szameit},\ and\ \citenamefont
		{Kivshar}}]{Grafe12}%
	\BibitemOpen
	\bibfield  {author} {\bibinfo {author} {\bibfnamefont {M.}~\bibnamefont
			{Gr{\"a}fe}}, \bibinfo {author} {\bibfnamefont {A.}~\bibnamefont {Solntsev}},
		\bibinfo {author} {\bibfnamefont {R.}~\bibnamefont {Keil}}, \bibinfo {author}
		{\bibfnamefont {A.}~\bibnamefont {Sukhorukov}}, \bibinfo {author}
		{\bibfnamefont {M.}~\bibnamefont {Heinrich}}, \bibinfo {author}
		{\bibfnamefont {A.}~\bibnamefont {T{\"u}nnermann}}, \bibinfo {author}
		{\bibfnamefont {S.}~\bibnamefont {Nolte}}, \bibinfo {author} {\bibfnamefont
			{A.}~\bibnamefont {Szameit}},\ and\ \bibinfo {author} {\bibfnamefont {Y.~S.}\
			\bibnamefont {Kivshar}},\ }\bibfield  {title} {\emph {\bibinfo {title}
			{{\color{blue}Biphoton generation in quadratic waveguide arrays: A classical
					optical simulation}}},\ }\href@noop {} {\bibfield  {journal} {\bibinfo
			{journal} {Scientific Reports}\ }\textbf {\bibinfo {volume} {2}} (\bibinfo
		{year} {2012})}\BibitemShut {NoStop}%
	\bibitem [{SM()}]{SM}%
	\BibitemOpen
\bibinfo {title} {\href@noop {} {}}\bibinfo {note} {See Supplemental Material for additional details on the experiment, theoretical framework, and numerical simulations.}\BibitemShut {Stop}%
	\bibitem [{\citenamefont {Su}\ \emph {et~al.}(1979)\citenamefont {Su},
		\citenamefont {Schrieffer},\ and\ \citenamefont {Heeger}}]{Su79}%
	\BibitemOpen
	\bibfield  {author} {\bibinfo {author} {\bibfnamefont {W.~P.}\ \bibnamefont
			{Su}}, \bibinfo {author} {\bibfnamefont {J.~R.}\ \bibnamefont {Schrieffer}},\
		and\ \bibinfo {author} {\bibfnamefont {A.~J.}\ \bibnamefont {Heeger}},\
	}\bibfield  {title} {\emph {\bibinfo {title} {{\color{blue}Solitons in
					Polyacetylene}}},\ }\href@noop {} {\bibfield  {journal} {\bibinfo  {journal}
			{Phys. Rev. Lett.}\ }\textbf {\bibinfo {volume} {42}},\ \bibinfo {pages}
		{1698} (\bibinfo {year} {1979})}\BibitemShut {NoStop}%
	\bibitem [{\citenamefont {Blanco-Redondo}\ \emph {et~al.}(2016)\citenamefont
		{Blanco-Redondo}, \citenamefont {Andonegui}, \citenamefont {Collins},
		\citenamefont {Harari}, \citenamefont {Lumer}, \citenamefont {Rechtsman},
		\citenamefont {Eggleton},\ and\ \citenamefont {Segev}}]{BlancoRedondo16}%
	\BibitemOpen
	\bibfield  {author} {\bibinfo {author} {\bibfnamefont {A.}~\bibnamefont
			{Blanco-Redondo}}, \bibinfo {author} {\bibfnamefont {I.}~\bibnamefont
			{Andonegui}}, \bibinfo {author} {\bibfnamefont {M.~J.}\ \bibnamefont
			{Collins}}, \bibinfo {author} {\bibfnamefont {G.}~\bibnamefont {Harari}},
		\bibinfo {author} {\bibfnamefont {Y.}~\bibnamefont {Lumer}}, \bibinfo
		{author} {\bibfnamefont {M.~C.}\ \bibnamefont {Rechtsman}}, \bibinfo {author}
		{\bibfnamefont {B.~J.}\ \bibnamefont {Eggleton}},\ and\ \bibinfo {author}
		{\bibfnamefont {M.}~\bibnamefont {Segev}},\ }\bibfield  {title} {\emph
		{\bibinfo {title} {{\color{blue}Topological optical waveguiding in silicon
					and the transition between topological and trivial defect states}}},\
	}\href@noop {} {\bibfield  {journal} {\bibinfo  {journal} {Phys. Rev. Lett.}\
		}\textbf {\bibinfo {volume} {116}},\ \bibinfo {pages} {163901} (\bibinfo
		{year} {2016})}\BibitemShut {NoStop}%
	\bibitem [{\citenamefont {Asb{\'o}th}\ \emph {et~al.}(2016)\citenamefont
		{Asb{\'o}th}, \citenamefont {Oroszl{\'a}ny},\ and\ \citenamefont
		{P{\'a}lyi}}]{Asboth16}%
	\BibitemOpen
	\bibfield  {author} {\bibinfo {author} {\bibfnamefont {J.~K.}\ \bibnamefont
			{Asb{\'o}th}}, \bibinfo {author} {\bibfnamefont {L.}~\bibnamefont
			{Oroszl{\'a}ny}},\ and\ \bibinfo {author} {\bibfnamefont {A.}~\bibnamefont
			{P{\'a}lyi}},\ }\href@noop {} {\emph {\bibinfo {title} {A short course on
				topological insulators}}},\ Vol.\ \bibinfo {volume} {919}\ (\bibinfo
	{publisher} {Springer},\ \bibinfo {year} {2016})\BibitemShut {NoStop}%
	\bibitem [{\citenamefont {St-Jean}\ \emph {et~al.}(2017)\citenamefont
		{St-Jean}, \citenamefont {Goblot}, \citenamefont {Galopin}, \citenamefont
		{Lema{\^\i}tre}, \citenamefont {Ozawa}, \citenamefont {Le~Gratiet},
		\citenamefont {Sagnes}, \citenamefont {Bloch},\ and\ \citenamefont
		{Amo}}]{StJean17}%
	\BibitemOpen
	\bibfield  {author} {\bibinfo {author} {\bibfnamefont {P.}~\bibnamefont
			{St-Jean}}, \bibinfo {author} {\bibfnamefont {V.}~\bibnamefont {Goblot}},
		\bibinfo {author} {\bibfnamefont {E.}~\bibnamefont {Galopin}}, \bibinfo
		{author} {\bibfnamefont {A.}~\bibnamefont {Lema{\^\i}tre}}, \bibinfo {author}
		{\bibfnamefont {T.}~\bibnamefont {Ozawa}}, \bibinfo {author} {\bibfnamefont
			{L.}~\bibnamefont {Le~Gratiet}}, \bibinfo {author} {\bibfnamefont
			{I.}~\bibnamefont {Sagnes}}, \bibinfo {author} {\bibfnamefont
			{J.}~\bibnamefont {Bloch}},\ and\ \bibinfo {author} {\bibfnamefont
			{A.}~\bibnamefont {Amo}},\ }\bibfield  {title} {\emph {\bibinfo {title}
			{{\color{blue}Lasing in topological edge states of a one-dimensional
					lattice}}},\ }\href@noop {} {\bibfield  {journal} {\bibinfo  {journal}
			{Nature Photonics}\ }\textbf {\bibinfo {volume} {11}},\ \bibinfo {pages}
		{651} (\bibinfo {year} {2017})}\BibitemShut {NoStop}%
	\bibitem [{\citenamefont {Baboux}\ \emph {et~al.}(2015)\citenamefont {Baboux},
		\citenamefont {Perez}, \citenamefont {Ullrich}, \citenamefont {Karczewski},\
		and\ \citenamefont {Wojtowicz}}]{Kailo15}%
	\BibitemOpen
	\bibfield  {author} {\bibinfo {author} {\bibfnamefont {F.}~\bibnamefont
			{Baboux}}, \bibinfo {author} {\bibfnamefont {F.}~\bibnamefont {Perez}},
		\bibinfo {author} {\bibfnamefont {C.~A.}\ \bibnamefont {Ullrich}}, \bibinfo
		{author} {\bibfnamefont {G.}~\bibnamefont {Karczewski}},\ and\ \bibinfo
		{author} {\bibfnamefont {T.}~\bibnamefont {Wojtowicz}},\ }\bibfield  {title}
	{\emph {\bibinfo {title} {{\color{blue}Electron density magnification of the
					collective spin-orbit field in quantum wells}}},\ }\href@noop {} {\bibfield
		{journal} {\bibinfo  {journal} {Phys. Rev. B}\ }\textbf {\bibinfo {volume}
			{92}},\ \bibinfo {pages} {125307} (\bibinfo {year} {2015})}\BibitemShut
	{NoStop}%
	\bibitem [{\citenamefont {Dietrich}\ \emph {et~al.}(2016)\citenamefont
		{Dietrich}, \citenamefont {Fiore}, \citenamefont {Thompson}, \citenamefont
		{Kamp},\ and\ \citenamefont {H{\"o}fling}}]{Dietrich16}%
	\BibitemOpen
	\bibfield  {author} {\bibinfo {author} {\bibfnamefont {C.~P.}\ \bibnamefont
			{Dietrich}}, \bibinfo {author} {\bibfnamefont {A.}~\bibnamefont {Fiore}},
		\bibinfo {author} {\bibfnamefont {M.~G.}\ \bibnamefont {Thompson}}, \bibinfo
		{author} {\bibfnamefont {M.}~\bibnamefont {Kamp}},\ and\ \bibinfo {author}
		{\bibfnamefont {S.}~\bibnamefont {H{\"o}fling}},\ }\bibfield  {title} {\emph
		{\bibinfo {title} {{\color{blue}{GaAs} integrated quantum photonics: Towards
					compact and multi-functional quantum photonic integrated circuits}}},\
	}\href@noop {} {\bibfield  {journal} {\bibinfo  {journal} {Laser \& Photonics
				Reviews}\ }\textbf {\bibinfo {volume} {10}},\ \bibinfo {pages} {870}
		(\bibinfo {year} {2016})}\BibitemShut {NoStop}%
	\bibitem [{\citenamefont {Helmy}\ \emph {et~al.}(2011)\citenamefont {Helmy},
		\citenamefont {Abolghasem}, \citenamefont {Stewart~Aitchison}, \citenamefont
		{Bijlani}, \citenamefont {Han}, \citenamefont {Holmes}, \citenamefont
		{Hutchings}, \citenamefont {Younis},\ and\ \citenamefont {Wagner}}]{Helmy11}%
	\BibitemOpen
	\bibfield  {author} {\bibinfo {author} {\bibfnamefont {A.~S.}\ \bibnamefont
			{Helmy}}, \bibinfo {author} {\bibfnamefont {P.}~\bibnamefont {Abolghasem}},
		\bibinfo {author} {\bibfnamefont {J.}~\bibnamefont {Stewart~Aitchison}},
		\bibinfo {author} {\bibfnamefont {B.~J.}\ \bibnamefont {Bijlani}}, \bibinfo
		{author} {\bibfnamefont {J.}~\bibnamefont {Han}}, \bibinfo {author}
		{\bibfnamefont {B.~M.}\ \bibnamefont {Holmes}}, \bibinfo {author}
		{\bibfnamefont {D.~C.}\ \bibnamefont {Hutchings}}, \bibinfo {author}
		{\bibfnamefont {U.}~\bibnamefont {Younis}},\ and\ \bibinfo {author}
		{\bibfnamefont {S.~J.}\ \bibnamefont {Wagner}},\ }\bibfield  {title} {\emph
		{\bibinfo {title} {{\color{blue}Recent advances in phase matching of
					second-order nonlinearities in monolithic semiconductor waveguides}}},\
	}\href@noop {} {\bibfield  {journal} {\bibinfo  {journal} {Laser \& Photonics
				Reviews}\ }\textbf {\bibinfo {volume} {5}},\ \bibinfo {pages} {272} (\bibinfo
		{year} {2011})}\BibitemShut {NoStop}%
	\bibitem [{\citenamefont {Horn}\ \emph {et~al.}(2012)\citenamefont {Horn},
		\citenamefont {Abolghasem}, \citenamefont {Bijlani}, \citenamefont {Kang},
		\citenamefont {Helmy},\ and\ \citenamefont {Weihs}}]{Horn12}%
	\BibitemOpen
	\bibfield  {author} {\bibinfo {author} {\bibfnamefont {R.}~\bibnamefont
			{Horn}}, \bibinfo {author} {\bibfnamefont {P.}~\bibnamefont {Abolghasem}},
		\bibinfo {author} {\bibfnamefont {B.~J.}\ \bibnamefont {Bijlani}}, \bibinfo
		{author} {\bibfnamefont {D.}~\bibnamefont {Kang}}, \bibinfo {author}
		{\bibfnamefont {A.~S.}\ \bibnamefont {Helmy}},\ and\ \bibinfo {author}
		{\bibfnamefont {G.}~\bibnamefont {Weihs}},\ }\bibfield  {title} {\emph
		{\bibinfo {title} {{\color{blue}Monolithic Source of Photon Pairs}}},\
	}\href@noop {} {\bibfield  {journal} {\bibinfo  {journal} {Phys. Rev. Lett.}\
		}\textbf {\bibinfo {volume} {108}},\ \bibinfo {pages} {153605} (\bibinfo
		{year} {2012})}\BibitemShut {NoStop}%
	\bibitem [{\citenamefont {Baboux}\ \emph {et~al.}(2023)\citenamefont {Baboux},
		\citenamefont {Moody},\ and\ \citenamefont {Ducci}}]{Baboux23}%
	\BibitemOpen
	\bibfield  {author} {\bibinfo {author} {\bibfnamefont {F.}~\bibnamefont
			{Baboux}}, \bibinfo {author} {\bibfnamefont {G.}~\bibnamefont {Moody}},\ and\
		\bibinfo {author} {\bibfnamefont {S.}~\bibnamefont {Ducci}},\ }\bibfield
	{title} {\emph {\bibinfo {title} {{\color{blue}Nonlinear integrated quantum
					photonics with {AlGaAs}}}},\ }\href@noop {} {\bibfield  {journal} {\bibinfo
			{journal} {Optica}\ }\textbf {\bibinfo {volume} {10}},\ \bibinfo {pages}
		{917} (\bibinfo {year} {2023})}\BibitemShut {NoStop}%
	\bibitem [{\citenamefont {Spring}\ \emph {et~al.}(2017)\citenamefont {Spring},
		\citenamefont {Mennea}, \citenamefont {Metcalf}, \citenamefont {Humphreys},
		\citenamefont {Gates}, \citenamefont {Rogers}, \citenamefont {S{\"o}ller},
		\citenamefont {Smith}, \citenamefont {Kolthammer}, \citenamefont {Smith}
		\emph {et~al.}}]{Spring17}%
	\BibitemOpen
	\bibfield  {author} {\bibinfo {author} {\bibfnamefont {J.~B.}\ \bibnamefont
			{Spring}}, \bibinfo {author} {\bibfnamefont {P.~L.}\ \bibnamefont {Mennea}},
		\bibinfo {author} {\bibfnamefont {B.~J.}\ \bibnamefont {Metcalf}}, \bibinfo
		{author} {\bibfnamefont {P.~C.}\ \bibnamefont {Humphreys}}, \bibinfo {author}
		{\bibfnamefont {J.~C.}\ \bibnamefont {Gates}}, \bibinfo {author}
		{\bibfnamefont {H.~L.}\ \bibnamefont {Rogers}}, \bibinfo {author}
		{\bibfnamefont {C.}~\bibnamefont {S{\"o}ller}}, \bibinfo {author}
		{\bibfnamefont {B.~J.}\ \bibnamefont {Smith}}, \bibinfo {author}
		{\bibfnamefont {W.~S.}\ \bibnamefont {Kolthammer}}, \bibinfo {author}
		{\bibfnamefont {P.~G.}\ \bibnamefont {Smith}}, \emph {et~al.},\ }\bibfield
	{title} {\emph {\bibinfo {title} {{\color{blue}Chip-based array of
					near-identical, pure, heralded single-photon sources}}},\ }\href@noop {}
	{\bibfield  {journal} {\bibinfo  {journal} {Optica}\ }\textbf {\bibinfo
			{volume} {4}},\ \bibinfo {pages} {90} (\bibinfo {year} {2017})}\BibitemShut
	{NoStop}%
	\bibitem [{\citenamefont {Brod}\ \emph {et~al.}(2019)\citenamefont {Brod},
		\citenamefont {Galv{\~a}o}, \citenamefont {Crespi}, \citenamefont {Osellame},
		\citenamefont {Spagnolo},\ and\ \citenamefont {Sciarrino}}]{Brod19}%
	\BibitemOpen
	\bibfield  {author} {\bibinfo {author} {\bibfnamefont {D.~J.}\ \bibnamefont
			{Brod}}, \bibinfo {author} {\bibfnamefont {E.~F.}\ \bibnamefont
			{Galv{\~a}o}}, \bibinfo {author} {\bibfnamefont {A.}~\bibnamefont {Crespi}},
		\bibinfo {author} {\bibfnamefont {R.}~\bibnamefont {Osellame}}, \bibinfo
		{author} {\bibfnamefont {N.}~\bibnamefont {Spagnolo}},\ and\ \bibinfo
		{author} {\bibfnamefont {F.}~\bibnamefont {Sciarrino}},\ }\bibfield  {title}
	{\emph {\bibinfo {title} {{\color{blue}Photonic implementation of boson
					sampling: a review}}},\ }\href@noop {} {\bibfield  {journal} {\bibinfo
			{journal} {Advanced Photonics}\ }\textbf {\bibinfo {volume} {1}},\ \bibinfo
		{pages} {034001} (\bibinfo {year} {2019})}\BibitemShut {NoStop}%
	\bibitem [{\citenamefont {Chen}\ \emph {et~al.}(2007)\citenamefont {Chen},
		\citenamefont {Lee},\ and\ \citenamefont {Kumar}}]{Chen07}%
	\BibitemOpen
	\bibfield  {author} {\bibinfo {author} {\bibfnamefont {J.}~\bibnamefont
			{Chen}}, \bibinfo {author} {\bibfnamefont {K.~F.}\ \bibnamefont {Lee}},\ and\
		\bibinfo {author} {\bibfnamefont {P.}~\bibnamefont {Kumar}},\ }\bibfield
	{title} {\emph {\bibinfo {title} {{\color{blue}Deterministic quantum splitter
					based on time-reversed Hong-Ou-Mandel interference}}},\ }\href@noop {}
	{\bibfield  {journal} {\bibinfo  {journal} {Phys. Rev. A}\ }\textbf {\bibinfo
			{volume} {76}},\ \bibinfo {pages} {031804} (\bibinfo {year}
		{2007})}\BibitemShut {NoStop}%
	\bibitem [{\citenamefont {Fabre}\ \emph {et~al.}(2020)\citenamefont {Fabre},
		\citenamefont {Belhassen}, \citenamefont {Minneci},
		\citenamefont {Felicetti}, \citenamefont {Keller},
		\citenamefont {Amanti}, \citenamefont {Baboux},
		\citenamefont {Coudreau}, \citenamefont {Ducci},
		\citenamefont {Milman}}]{Fabre20}%
	\BibitemOpen
	\bibfield {author} {
		\bibinfo {author} {\bibfnamefont {N.}~\bibnamefont {Fabre}},
		\bibinfo {author} {\bibfnamefont {J.}~\bibnamefont {Belhassen}},
		\bibinfo {author} {\bibfnamefont {A.}~\bibnamefont {Minneci}},
		\bibinfo {author} {\bibfnamefont {S.}~\bibnamefont {Felicetti}},
		\bibinfo {author} {\bibfnamefont {A.}~\bibnamefont {Keller}},
		\bibinfo {author} {\bibfnamefont {M.~I.}~\bibnamefont {Amanti}},
		\bibinfo {author} {\bibfnamefont {F.}~\bibnamefont {Baboux}},
		\bibinfo {author} {\bibfnamefont {T.}~\bibnamefont {Coudreau}},
		\bibinfo {author} {\bibfnamefont {S.}~\bibnamefont {Ducci}},
		\bibinfo {author} {\bibfnamefont {P.}~\bibnamefont {Milman}},\
	}
	\bibfield {title} {\emph {
			\bibinfo {title} {{\color{blue}Producing a delocalized
					frequency-time Schrödinger-cat-like state with
					Hong-Ou-Mandel interferometry}}},\
	}
	\href@noop {} {
		\bibfield {journal} {\bibinfo {journal} {Phys. Rev. A}\ }
		\textbf {\bibinfo {volume} {102}},\
		\bibinfo {pages} {023710}
		(\bibinfo {year} {2020})
	}\BibitemShut {NoStop}%
	\bibitem [{\citenamefont {Fabre}\ \emph {et~al.}(2022)\citenamefont {Fabre},
		\citenamefont {Amanti}, \citenamefont {Baboux}, \citenamefont {Keller},
		\citenamefont {Ducci},\ and\ \citenamefont {Milman}}]{Fabre22}%
	\BibitemOpen
	\bibfield  {author} {\bibinfo {author} {\bibfnamefont {N.}~\bibnamefont
			{Fabre}}, \bibinfo {author} {\bibfnamefont {M.}~\bibnamefont {Amanti}},
		\bibinfo {author} {\bibfnamefont {F.}~\bibnamefont {Baboux}}, \bibinfo
		{author} {\bibfnamefont {A.}~\bibnamefont {Keller}}, \bibinfo {author}
		{\bibfnamefont {S.}~\bibnamefont {Ducci}},\ and\ \bibinfo {author}
		{\bibfnamefont {P.}~\bibnamefont {Milman}},\ }\bibfield  {title} {\emph
		{\bibinfo {title} {{\color{blue}The {Hong--Ou--Mandel} experiment: from
					photon indistinguishability to continuous-variable quantum computing}}},\
	}\href@noop {} {\bibfield  {journal} {\bibinfo  {journal} {The European
				Physical Journal D}\ }\textbf {\bibinfo {volume} {76}},\ \bibinfo {pages}
		{196} (\bibinfo {year} {2022})}\BibitemShut {NoStop}%
	\bibitem [{\citenamefont {Meskine}\ \emph {et~al.}(2024)\citenamefont
		{Meskine}, \citenamefont {Descamps}, \citenamefont {Keller}, \citenamefont
		{Lema{\^\i}tre}, \citenamefont {Baboux}, \citenamefont {Ducci},\ and\
		\citenamefont {Milman}}]{Meskine24}%
	\BibitemOpen
	\bibfield  {author} {\bibinfo {author} {\bibfnamefont {O.}~\bibnamefont
			{Meskine}}, \bibinfo {author} {\bibfnamefont {E.}~\bibnamefont {Descamps}},
		\bibinfo {author} {\bibfnamefont {A.}~\bibnamefont {Keller}}, \bibinfo
		{author} {\bibfnamefont {A.}~\bibnamefont {Lema{\^\i}tre}}, \bibinfo {author}
		{\bibfnamefont {F.}~\bibnamefont {Baboux}}, \bibinfo {author} {\bibfnamefont
			{S.}~\bibnamefont {Ducci}},\ and\ \bibinfo {author} {\bibfnamefont
			{P.}~\bibnamefont {Milman}},\ }\bibfield  {title} {\emph {\bibinfo {title}
			{{\color{blue}Approaching maximal precision of Hong-Ou-Mandel interferometry
					with nonperfect visibility}}},\ }\href@noop {} {\bibfield  {journal}
		{\bibinfo  {journal} {Physical Review Letters}\ }\textbf {\bibinfo {volume}
			{132}},\ \bibinfo {pages} {193603} (\bibinfo {year} {2024})}\BibitemShut
	{NoStop}%
	\bibitem [{\citenamefont {Descamps}\ and\ \citenamefont
		{Milman}(2026)}]{Descamps26}%
	\BibitemOpen
	\bibfield  {author} {\bibinfo {author} {\bibfnamefont {{\'E}.}~\bibnamefont
			{Descamps}}\ and\ \bibinfo {author} {\bibfnamefont {P.}~\bibnamefont
			{Milman}},\ }\bibfield  {title} {\emph {\bibinfo {title}
			{{\color{blue}Spatial symmetry in few-photon interference and quantum
					metrology}}},\ }\href@noop {} {\bibfield  {journal} {\bibinfo  {journal}
			{arXiv:2609.12786}\ } (\bibinfo {year} {2026})}\BibitemShut {NoStop}%
	\bibitem [{\citenamefont {Scollon}\ and\ \citenamefont
		{Kennett}(2020)}]{Scollon20}%
	\BibitemOpen
	\bibfield  {author} {\bibinfo {author} {\bibfnamefont {M.}~\bibnamefont
			{Scollon}}\ and\ \bibinfo {author} {\bibfnamefont {M.~P.}\ \bibnamefont
			{Kennett}},\ }\bibfield  {title} {\emph {\bibinfo {title}
			{{\color{blue}Persistence of chirality in the Su-Schrieffer-Heeger model in
					the presence of on-site disorder}}},\ }\href@noop {} {\bibfield  {journal}
		{\bibinfo  {journal} {Physical Review B}\ }\textbf {\bibinfo {volume}
			{101}},\ \bibinfo {pages} {144204} (\bibinfo {year} {2020})}\BibitemShut
	{NoStop}%
	\bibitem [{\citenamefont {Bergamasco}\ \emph {et~al.}(2021)\citenamefont
		{Bergamasco}, \citenamefont {Sipe},\ and\ \citenamefont
		{Liscidini}}]{Bergamasco21}%
	\BibitemOpen
	\bibfield  {author} {\bibinfo {author} {\bibfnamefont {N.}~\bibnamefont
			{Bergamasco}}, \bibinfo {author} {\bibfnamefont {J.}~\bibnamefont {Sipe}},\
		and\ \bibinfo {author} {\bibfnamefont {M.}~\bibnamefont {Liscidini}},\
	}\bibfield  {title} {\emph {\bibinfo {title} {{\color{blue}Generation of
					hyper-entangled states in strongly coupled topological defects}}},\
	}\href@noop {} {\bibfield  {journal} {\bibinfo  {journal} {Optics Lett.}\
		}\textbf {\bibinfo {volume} {46}},\ \bibinfo {pages} {2244} (\bibinfo {year}
		{2021})}\BibitemShut {NoStop}%
	\bibitem [{\citenamefont {Elshaari}\ \emph {et~al.}(2020)\citenamefont
		{Elshaari}, \citenamefont {Pernice}, \citenamefont {Srinivasan},
		\citenamefont {Benson},\ and\ \citenamefont {Zwiller}}]{Elshaari20}%
	\BibitemOpen
	\bibfield  {author} {\bibinfo {author} {\bibfnamefont {A.~W.}\ \bibnamefont
			{Elshaari}}, \bibinfo {author} {\bibfnamefont {W.}~\bibnamefont {Pernice}},
		\bibinfo {author} {\bibfnamefont {K.}~\bibnamefont {Srinivasan}}, \bibinfo
		{author} {\bibfnamefont {O.}~\bibnamefont {Benson}},\ and\ \bibinfo {author}
		{\bibfnamefont {V.}~\bibnamefont {Zwiller}},\ }\bibfield  {title} {\emph
		{\bibinfo {title} {{\color{blue}Hybrid integrated quantum photonic
					circuits}}},\ }\href@noop {} {\bibfield  {journal} {\bibinfo  {journal}
			{Nature Photonics}\ }\textbf {\bibinfo {volume} {14}},\ \bibinfo {pages}
		{285} (\bibinfo {year} {2020})}\BibitemShut {NoStop}%
	\bibitem [{\citenamefont {Yuan}\ \emph {et~al.}(2018)\citenamefont {Yuan},
		\citenamefont {Lin}, \citenamefont {Xiao},\ and\ \citenamefont
		{Fan}}]{Yuan18}%
	\BibitemOpen
	\bibfield  {author} {\bibinfo {author} {\bibfnamefont {L.}~\bibnamefont
			{Yuan}}, \bibinfo {author} {\bibfnamefont {Q.}~\bibnamefont {Lin}}, \bibinfo
		{author} {\bibfnamefont {M.}~\bibnamefont {Xiao}},\ and\ \bibinfo {author}
		{\bibfnamefont {S.}~\bibnamefont {Fan}},\ }\bibfield  {title} {\emph
		{\bibinfo {title} {{\color{blue}Synthetic dimension in photonics}}},\
	}\href@noop {} {\bibfield  {journal} {\bibinfo  {journal} {Optica}\ }\textbf
		{\bibinfo {volume} {5}},\ \bibinfo {pages} {1396} (\bibinfo {year}
		{2018})}\BibitemShut {NoStop}%
	\bibitem [{\citenamefont {Wang}\ \emph {et~al.}(2014)\citenamefont {Wang},
		\citenamefont {Santamato}, \citenamefont {Jiang}, \citenamefont {Bonneau},
		\citenamefont {Engin}, \citenamefont {Silverstone}, \citenamefont {Lermer},
		\citenamefont {Beetz}, \citenamefont {Kamp}, \citenamefont {H{\"o}fling}
		\emph {et~al.}}]{Wang14}%
	\BibitemOpen
	\bibfield  {author} {\bibinfo {author} {\bibfnamefont {J.}~\bibnamefont
			{Wang}}, \bibinfo {author} {\bibfnamefont {A.}~\bibnamefont {Santamato}},
		\bibinfo {author} {\bibfnamefont {P.}~\bibnamefont {Jiang}}, \bibinfo
		{author} {\bibfnamefont {D.}~\bibnamefont {Bonneau}}, \bibinfo {author}
		{\bibfnamefont {E.}~\bibnamefont {Engin}}, \bibinfo {author} {\bibfnamefont
			{J.~W.}\ \bibnamefont {Silverstone}}, \bibinfo {author} {\bibfnamefont
			{M.}~\bibnamefont {Lermer}}, \bibinfo {author} {\bibfnamefont
			{J.}~\bibnamefont {Beetz}}, \bibinfo {author} {\bibfnamefont
			{M.}~\bibnamefont {Kamp}}, \bibinfo {author} {\bibfnamefont {S.}~\bibnamefont
			{H{\"o}fling}}, \emph {et~al.},\ }\bibfield  {title} {\emph {\bibinfo {title}
			{{\color{blue}Gallium arsenide quantum photonic waveguide circuits}}},\
	}\href@noop {} {\bibfield  {journal} {\bibinfo  {journal} {Optics Comm.}\
		}\textbf {\bibinfo {volume} {327}},\ \bibinfo {pages} {49} (\bibinfo {year}
		{2014})}\BibitemShut {NoStop}%
	\bibitem [{\citenamefont {Piccioli}\ \emph {et~al.}(2022)\citenamefont
		{Piccioli}, \citenamefont {Szameit},\ and\ \citenamefont
		{Carusotto}}]{Piccioli22}%
	\BibitemOpen
	\bibfield  {author} {\bibinfo {author} {\bibfnamefont {F.~S.}\ \bibnamefont
			{Piccioli}}, \bibinfo {author} {\bibfnamefont {A.}~\bibnamefont {Szameit}},\
		and\ \bibinfo {author} {\bibfnamefont {I.}~\bibnamefont {Carusotto}},\
	}\bibfield  {title} {\emph {\bibinfo {title} {{\color{blue}Topologically
					protected frequency control of broadband signals in dynamically modulated
					waveguide arrays}}},\ }\href@noop {} {\bibfield  {journal} {\bibinfo
			{journal} {Phys. Rev. A}\ }\textbf {\bibinfo {volume} {105}},\ \bibinfo
		{pages} {053519} (\bibinfo {year} {2022})}\BibitemShut {NoStop}%
	\bibitem [{\citenamefont {Wu}\ \emph {et~al.}(2022)\citenamefont {Wu},
		\citenamefont {Wang}, \citenamefont {Li}, \citenamefont {Cheng},
		\citenamefont {Yu}, \citenamefont {Zheng}, \citenamefont {Yakovlev},
		\citenamefont {Yuan},\ and\ \citenamefont {Chen}}]{Wu22}%
	\BibitemOpen
	\bibfield  {author} {\bibinfo {author} {\bibfnamefont {X.}~\bibnamefont
			{Wu}}, \bibinfo {author} {\bibfnamefont {L.}~\bibnamefont {Wang}}, \bibinfo
		{author} {\bibfnamefont {G.}~\bibnamefont {Li}}, \bibinfo {author}
		{\bibfnamefont {D.}~\bibnamefont {Cheng}}, \bibinfo {author} {\bibfnamefont
			{D.}~\bibnamefont {Yu}}, \bibinfo {author} {\bibfnamefont {Y.}~\bibnamefont
			{Zheng}}, \bibinfo {author} {\bibfnamefont {V.~V.}\ \bibnamefont {Yakovlev}},
		\bibinfo {author} {\bibfnamefont {L.}~\bibnamefont {Yuan}},\ and\ \bibinfo
		{author} {\bibfnamefont {X.}~\bibnamefont {Chen}},\ }\bibfield  {title}
	{\emph {\bibinfo {title} {{\color{blue}Technologically feasible quasi-edge
					states and topological Bloch oscillation in the synthetic space}}},\
	}\href@noop {} {\bibfield  {journal} {\bibinfo  {journal} {Optics Express}\
		}\textbf {\bibinfo {volume} {30}},\ \bibinfo {pages} {24924} (\bibinfo {year}
		{2022})}\BibitemShut {NoStop}%
\end{thebibliography}
\end{document}